\documentclass[preprint,12pt,3p]{elsarticle}

\usepackage{fullpage}
\usepackage{graphicx}
\usepackage{subfigure}
\usepackage{amsfonts}
\usepackage{amsgen}
\usepackage{amsbsy}
\usepackage{amsmath}
\usepackage{amssymb}
\usepackage{mathrsfs}
\usepackage{mathtools}
\usepackage{bm}
\usepackage{epstopdf}
\usepackage{caption}
\usepackage{tabularx}
\usepackage{tabu}
\usepackage{xcolor}
\usepackage{nicefrac}
\usepackage{enumitem}
\usepackage[ruled]{algorithm2e}
\usepackage{url}

\newcommand{\Ge}[1]{\textcolor{black}{#1}}

\newcommand\etal{\textit{et al }}
\newcommand\ie{\textit{i.e. }}
\newcommand\viz{\textit{viz. }}
\newcommand\eg{\textit{e.g. }}
\newcommand{\e}[1]{\ensuremath{\times 10^{#1}}}
\newcommand\etc{\textit{etc}}
\DeclarePairedDelimiter\abs{\lvert}{\rvert}

\usepackage{amssymb}

\journal{Journal of Computational Physics}

\begin{document}

\begin{frontmatter}

\title{An efficient mass-preserving interface-correction level set/ghost fluid method for droplet suspensions under depletion forces}   

  \author[]{Zhouyang Ge\corref{cor1}}
  \cortext[cor1]{Corresponding author.}
  \ead{zhoge@mech.kth.se}
  
  \author[]{Jean-Christophe Loiseau\fnref{label2}}
  \fntext[label2]{Present address: Laboratoire DynFluid, Arts et M\'etiers ParisTech, 151 boulevard de l'h\^opital, 75013 Paris, France}
  \ead{jean-christophe.loiseau@ensam.eu}
  
  \author[]{Outi Tammisola}
  \ead{outi@mech.kth.se}
  
  \author[]{Luca Brandt}
  \ead{luca@mech.kth.se}

  \address{Linn\'e Flow Centre and SeRC (Swedish e-Science Research Centre), KTH Mechanics, \\ S-100 44 Stockholm, Sweden}

\sloppy
\begin{abstract}
Aiming for the simulation of colloidal droplets in microfluidic devices, we present here a numerical method for two-fluid systems subject to surface tension and depletion forces among the suspended droplets. The algorithm is based on \Ge{an efficient} solver for the incompressible two-phase Navier-Stokes equations, and uses a \Ge{mass-conserving} level set method to capture the fluid interface. 
The four novel ingredients proposed here are, firstly, an interface-correction level set (ICLS) method; 
global mass conservation is achieved by performing an additional advection near the interface, with a correction velocity obtained by locally solving an algebraic equation, which is easy to implement in both 2D and 3D. 
Secondly, we report a second-order accurate \Ge{geometric} estimation of the curvature at the interface and, 
thirdly, the combination of the ghost fluid method with the fast pressure-correction approach enabling an accurate and fast computation even for large density contrasts. 
Finally, we derive 
\Ge{a hydrodynamic model for the interaction forces}
induced by depletion of surfactant micelles and combine it with a multiple level set approach to study short-range interactions among droplets in the presence of attracting forces.
\end{abstract}


\end{frontmatter}




\section{Introduction}
\label{intro}

\begin{figure}[t!]
\centering
  \includegraphics[width=.8\columnwidth]{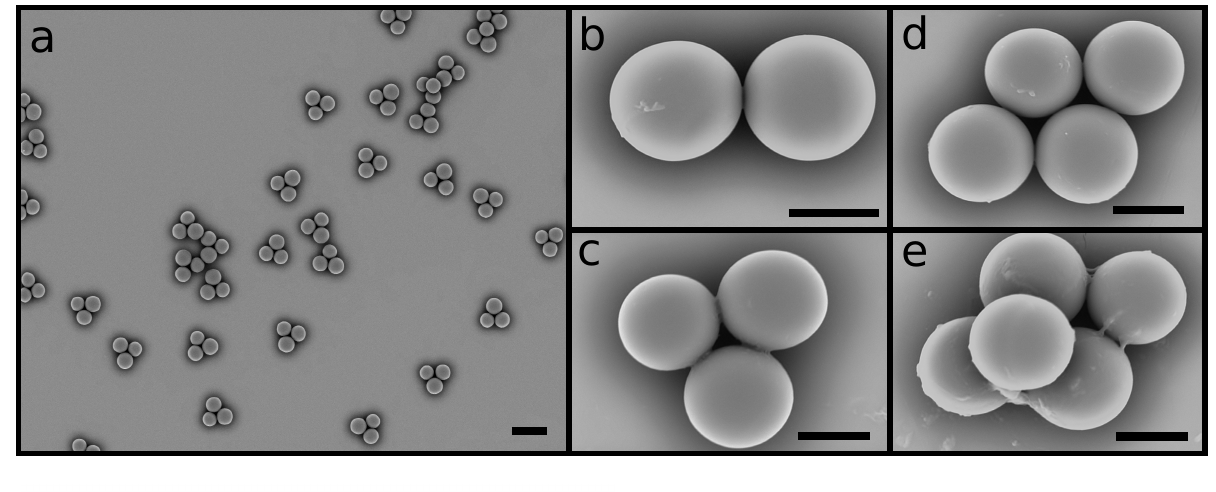}
  \caption{\Ge{Self-assembled colloidal clusters. a) Electron micrograph of a suspension of triplet clusters. Scale bar, 30 $\mu$m. b-e) Close up of doublet, triplet, quadruplet, and quintuplet clusters. Scale bars, 10 $\mu$m. Further details are available in \cite{Shen_AS_2016}, photograph courtesy of Dr.\ Joshua Ricouvier.}}
  \label{fig:exp}
\end{figure}

In the field of colloidal science, much progress has been made on the synthesis of elementary building blocks \Ge{(Fig.\ \ref{fig:exp})} mimicking molecular structures to elaborate innovative materials, \eg materials with complete three dimensional band gaps \cite{Xia_etal_AM2000, Velev_etal_AM2009, Li_etal_AC2011, Sacanna_etal_COCIS2011}. The basic elements of such colloidal molecules are particles or droplets less than one millimeter in size, and their self-assembly relies on either lengthy brownian motion or careful microfludic designs, on top of typical colloidal interactions, \eg depletion attraction and electrostatic repulsion \cite{Mewis_colloidal, Yi_etal_CM2013, Shen_AS_2016}. Regardless of the approach, however, questions remain why the colloidal particles/droplets undergo certain path to organize themselves and how such process can be controlled and optimized. Since full data are not yet accurately accessible from experiments in such miniature systems, computer simulations will be useful to provide supplemental information.

Scaling down to microscale appears first to be a convenience for the numerical simulations of multicomponent and multiphase systems as the non-linear Navier-Stokes (NS) equations can be reduced to the linear Stokes equations. This allows the use of boundary integral methods (BIM) \cite{Pozrikidis}, \eg most recently the GGEM-based BIM \cite{Kumar_JCP_2012, Lailai_SM_2014} solving the Stokes equations in general geometries. However, it is also possible to use the conventional unsteady, fractional-step/projection-method NS solver at low Reynolds number, combined with an interface description method \cite{Worner_2012, Galusinski_JCP_2008}. The latter approach is more versatile, probably less difficult to implement, and enjoys a rich literature of standard numerical techniques. Here, in view of a rich range of possible applications and considering also the rapid development of inertial microfluidics (where inertial effects are used to better control the flow behavior) we take the approach of simulating the incompressible, two-fluid NS as outlined in \cite{Dodd_JCP_2014}. The splitting procedure proposed in \cite{Dodd_JCP_2014} enables the use of fast solvers for the pressure Poisson equation also for large density and viscosity contrasts. The remaining choice then is to be made among the available interface-description methods. 

Generally, there are two categories of methods to resolve an interface in a NS solver, \ie front-tracking methods and front-capturing methods. An example of the front-tracking method is the immersed boundary method (IBM) \Ge{\cite{Peskin,Uhlmann}. Using Lagrangian points in a moving frame, IBM can offer a high interface resolution without the need to deform the underlying mesh in the fixed frame. However, the coupling of the two meshes relies on a regularized delta function, which introduces certain degrees of smearing. Moreover, large interface deformation requires frequent mesh rearrangement; and topology changes may have to be handled manually. These constraints make IBM typically more expensive and less appealing for droplet simulations.}

Front-capturing methods, on the other hand, are Eulerian and handle topology changes automatically; they are therefore easier to parallelize to achieve higher efficiency. One of such methods is the volume-of-fluid (VOF) method \cite{Scardovelli_ARFM_1999}, which defines different fluids with a discontinuous color function. The main advantage of VOF is its intrinsic mass conservation. It suffers however from inaccurate computations of the interface properties, \eg normals and curvatures. This makes it less favorable for simulations of microfluidic systems where surface tension \Ge{is the dominant effect and requires accurate modelling.}

Another popular front-capturing method is the level set (LS) method \cite{Sethian_levelset, Sussman_JCP_1994}. Contrary to VOF, LS prescribes the interface through a \Ge{(Lipschitz-)continuous} function which usually takes the form of the signed distance to the interface. \Ge{Under this definition}, normals and curvatures of the interface can be readily and accurately computed. However, the problem when simulating incompressible flows is that mass loss/gain may occur and accumulate because the LS function embeds no volume information. \Ge{In addition, errors can also arise from solving the LS advection equation and/or the reinitialization equation,} a procedure commonly required to reshape the LS into a distance function. Therefore, additional measures have to be taken to ensure mass conservation.

Many different approaches have been proposed to make LS mass-conserving, which can be classified into the following \Ge{four} methodologies. The first approach is to improve the LS discretization and reinitialization so that numerical errors are reduced. In practice, one can increase the order of LS fluxes \cite{Nourgaliev_JCP_2007}, minimize the displacement of the zero LS during reinitialization \Ge{\cite{Russo_JCP_2000,Nourgaliev_JCP_2007}}, or employ local mesh refinement \Ge{\cite{Strain_JCP_1999b,Min_JCP_2007,Herrmann_JCP_2008}}. By doing so, mass loss can be greatly reduced, although the LS function is still inherently non-conservative. The second remedy couples the LS with \Ge{a conservative description (\eg VOF) or Lagrangian particles.} For example, \Ge{the hybrid particle level set method \cite{Enright_JCP_2002},} the coupled level set volume-of-fluid (CLSVOF) method \cite{Sussman_JCP_2000}, the mass-conserving level set (MCLS) method \cite{Pijl_CVS_2008}, or the recent curvature-based mass-redistribution method \cite{Luo_JCP_2015}. With varying level of coupling, \Ge{these methods can usually preserve mass really well; the drawback is that the complexity and some inaccuracy (due to interpolation, reconstruction, \etc) of the other method will be imported. The third approach improves mass conservation by adding a volume-constraint in the LS or NS formulation. Examples of this kind include the interface-preserving LS redistancing algorithm \cite{Sussman_JSC_1997} and the mass-preserving NS projection method \cite{Salac_CPC_2016}. Finally, one can also smartly modify the definition of the LS, such as the hyperbolic-tangent level set \cite{Olsson_JCP_2005}, to reduce the overall mass loss.}

With the physical application of colloidal droplets in mind, 
\Ge{and using ideas from some of the above-mentioned methods, we heuristically propose an interface-correction level set (ICLS) method.} The essential idea of ICLS is to construct a normal velocity supported on the droplet interface and use it in an additional LS advection to compensate for mass loss, in a way similar to inflating a balloon. Because no coupling with VOF \Ge{or Lagrangian particles} is required, the simplicity and high accuracy of the original LS method is preserved, yet the extra computational cost of this procedure is negligible.


\Ge{Provided a mass-preserving level set method, the coupled flow solver must also accurately compute the surface tension, a singular effect of the normal stress on the interface. This is particularly important for microfluidic systems; as surface tension scales linearly with the dimension, it decays slower than volumetric forces (\eg gravity) when the size of the system reduces.}
To handle such discontinuities, one approach is the continuum surface force (CSF) \cite{Brackbill_JCP_1992}, \Ge{originally developed for the VOF method, later extended to the LS \cite{Sussman_JCP_1994}.} 
\Ge{Although easy to implement}, CSF effectively introduces an artificial spreading of the interface \Ge{by regularizing the pressure difference}, and it can become erroneous when two interfaces are within its smoothing width.
\Ge{A second, non-smearing approach is the ghost fluid method (GFM). Proposed initially for solving compressible Euler equations \cite{Fedkiw_JCP_1999}, GFM provides a finite-difference discretization of the gradient operator even if the stencil includes shocks. It has been proven to converge \cite{Liu_MC_2003} and was soon applied for treating the pressure jump in multiphase flows \cite{Kang_JSC_2000}. We note that although the GFM can be reformulated in a similar way to the CSF \cite{Lalanne_JCP_2015,Popinet_ARFM_2018}, its treatment for discontinuous quantities is sharp in the finite difference limit.} 

\Ge{Several implementation options of the GFM were suggested in \cite{Kang_JSC_2000,Lalanne_JCP_2015,Desjardins_JCP_2008}. Here, we follow the methodology of \cite{Desjardins_JCP_2008}, \ie using the GFM for the pressure jump due to surface tension while neglecting the viscous contribution. As will be discussed later, this choice is especially suitable for microfluidic applications where the capillary effect is strong.}
\Ge{To efficiently solve for the pressure, we further combine the GFM with a fast pressure-correction method (FastP*) \cite{Dodd_JCP_2014}. Such a combination enables a direct solve of the pressure Poisson equation using the Gauss elimination in the Fourier space; it is the most efficient when the computational domain is periodic, but it also applies to a range of homogeneous Dirichlet/Neumann boundary conditions via fast sine/cosine transforms \cite{Schumann_JCP_1988}, see \eg a recent open-source distribution \cite{Pedro_CaNS}. Using a second-order accurate, grid-converging interface curvature estimation, we will show that the coupled ICLS/NS solver can handle large density/viscosity contrasts and converges between first and second order in both space and time.}

Finally, a unique challenge to the simulation of colloidal droplets is the modeling of near-field interactions. It is known that two or more colloids can interact via dispersion, surface, depletion, and hydrodynamic forces \cite{Mewis_colloidal}. Apart from the hydrodynamic forces which is determined directly from the NS, and the dispersion forces which arise from quantum mechanical effects, the depletion and surface forces must be modelled. These forces can be either attraction or repulsion and are typically calculated from the gradient of a potential.
\Ge{Based on colloidal theory, we propose a novel hydrodynamic model for the depletion force in the framework of the ICLS/NS solver. Our method relies on two extensions: \textit{i)} extending the single level set (SLS) function to multiple level set (MLS) functions; and \textit{ii)} extending the GFM for computation of the gradient of depletion potential. 
MLS has the benefits that each droplet within a colloidal cluster can be treated individually, is allowed to interact with the other droplets, and is guarded from its own mass loss. MLS also prevents numerical coalescence of droplets when they get too close. The computational complexity, proportional to the number of MLS functions ($l$) and the number of cells in each dimension ($N$), is higher than SLS. However, we note that many techniques exist to reduce the CPU cost and/or memory consumption if $lN^d$ ($d=$ 2 or 3) is large. For detailed implementations of such optimized algorithms we refer to \cite{Peng_JCP_1999,Nielsen_JSC_2006,Brun_JCP_2012}. In the present paper, we will demonstrate the self-assembly of colloidal droplets using one droplet per MLS function.}

The paper is organized as follows. In Sec.\ \ref{subsec: gov eqns}, the governing equations for the incompressible, two-phase flow are briefly presented. In Sec.\ \ref{subsec: dls}, the classical signed-distance LS methodology \Ge{together with some commonly used numerical schemes is discussed.} \Ge{We then introduce the ICLS method in Sec.\ \ref{subsec: ICLS}, starting from the derivation ending with a demonstration. We further provide a geometric estimation of the interface curvature tailored to the GFM in Sec.\ \ref{subsec: curv}. The complete ICLS/NS solver is outlined in Sec.\ \ref{subsec: NS}, including a detailed description of the implementation and three examples of validation.} In Sec.\ \ref{sec: idrop}, we propose a MLS/GFM-based method for the modeling of near-field depletion potential. \Ge{Finally, we summarize the overall methodology in Sec.\ \ref{sec: conclusion}.}


\section{Governing equations for interfacial two-phase flow}
\label{subsec: gov eqns}

The dynamics of the incompressible flow of two immiscible fluids is governed by the Navier-Stokes equations, written in the non-dimensional form

\begin{subequations}
 \begin{equation}
   \nabla \cdot {\bm u} = 0,
  \label{div free}
 \end{equation}

 \begin{equation}
      \frac{\partial {\bm u}}{\partial t} + {\bm u} \cdot \nabla {\bm u} = \frac{1}{\rho_i} \bigg(-\nabla p + \frac{1}{Re} \nabla \cdot \big[ \mu_i ( \nabla {\bm u} + \nabla {\bm u}^T ) \big] \bigg) + \frac{1}{Fr}{\bm g},
  \label{NS}
 \end{equation}
\end{subequations}

\noindent where ${\bm u}={\bm u}({\bm x},t)$ is the velocity field, $p=p({\bm x},t)$ is the pressure field, and $\bm{g}$ is a unit vector aligned with gravity or buoyancy. $\rho_i$ and $\mu_i$ are the density and dynamic viscosity ratios of fluid $i$ ($i=1$ or $2$) and the reference fluid. These properties are \Ge{constant in each phase and} subject to a jump across the interface, which we denote as $[\rho]_\Gamma=\rho_2-\rho_1$ for density and $[\mu]_\Gamma=\mu_2-\mu_1$ for viscosity. For viscous flows, the velocity and its tangential derivatives are continuous on the interface \cite{Liu_JCP_1994}. However, the pressure is discontinuous due to the surface tension and the viscosity jump, \ie
\begin{equation}
    [p]_\Gamma = \frac{1}{We} \kappa + \frac{2}{Re}[\mu]_\Gamma {\bm n}^T \cdot \nabla {\bm u} \cdot {\bm n},
  \label{pressure jump}
\end{equation}
\noindent where $\kappa$ is the interface curvature, and ${\bm n}$ is the normal to the interface.  
If the surface tension coefficient, $\tilde{\sigma}$, varies on the interface the tangential stress is also discontinuous. In this paper, we assume constant and uniform $\tilde{\sigma}$. In Eqs.\ \eqref{NS} and \eqref{pressure jump}, Re, We, and Fr are, respectively, the Reynolds, Weber, and Froude numbers, defined as
\begin{equation}
  \begin{aligned}
    Re = \frac{\tilde{\rho_1} \tilde{U} \tilde{L}}{\tilde{\mu_1}},\quad \quad We = \frac{\tilde{\rho_1} \tilde{U}^2 \tilde{L}}{\tilde{\sigma}},\quad \quad Fr=\frac{\tilde{U}^2}{\tilde{g}\tilde{L}},      
  \label{non-di}    
  \end{aligned}
\end{equation}
\noindent where $\tilde{U}$, $\tilde{L}$, $\tilde{\rho_1}$, $\tilde{\mu_1}$, and $\tilde{g}$ denote the reference dimensional velocity, length, density, dynamic viscosity, and gravitational acceleration. Note that $\rho_1=1$ and $\mu_1=1$ (i.e.\ we define fluid 1 as the reference fluid).


\section{Classical level set methodology}
\label{subsec: dls}

In the level set framework, the interface $\Gamma$ is defined implicitly as the zero value of a scalar function $\phi({\bm x},t)$, \ie $\Gamma = \{ {\bm x} ~ \rvert ~ \phi({\bm x},t) = 0 \}$. Mathematically, $\phi({\bm x},t)$ can be any smooth or non-smooth function; but it is classically shaped as the signed Euclidean distance to the interface \cite{Mulder_JCP_1992, Sussman_JCP_1994}, \viz

\begin{equation}
    \phi({\bm x},t) = sgn({\bm x}) |{\bm x}-{\bm x_\Gamma}|,
  \label{dist ls}
\end{equation}

 \noindent where ${\bm x}_\Gamma$ denotes the closest point on the interface from nodal point ${\bm x}$, and $sgn({\bm x})$ is a sign function equal to $1$ or $-1$ depending on which side of the interface it lies. For two-phase problems with single level set, $sgn({\bm x})$ provides a natural ``color function" for phase indication. Furthermore, with this definition, geometric properties such as the unit normal vector, ${\bm n}$, and the local mean curvature, $\kappa$, can be conveniently computed as

\begin{equation}
    {\bm n} = \frac{\nabla \phi}{\abs{ \nabla \phi }},
  \label{normal}
\end{equation}

\begin{equation}
    \kappa = -\nabla \cdot {\bm n}.
  \label{curv}
\end{equation}


\subsection{Advection}
\label{ssubsec: ls adv}

The motion of a fluid interface is governed by the following PDE
\begin{equation}
  \frac{\partial \phi}{\partial t} + {\bm u} \cdot \nabla \phi = 0,
  \label{ls adv}
\end{equation}
where ${\bm u}$ is the flow velocity field. Despite of its simple form, obtaining an accurate and robust solution to Eq.\ \eqref{ls adv} is challenging. For two-fluid problems, state-of-the-art level set transport schemes include the high-order upstream-central (HOUC) scheme \cite{Nourgaliev_JCP_2007}, the weighted essentially non-oscillatory (WENO) scheme \cite{Liu_JCP_1994}, \Ge{the semi-Lagrangian scheme \cite{Strain_JCP_1999}}, or the semi-jet scheme \cite{Velmurugana_AX_2016}. Quantitative comparisons of these schemes in various test cases can be found in \cite{Nourgaliev_JCP_2007, Velmurugana_AX_2016}. We note that the choice of the scheme is case-dependent, \ie depending on the smoothness of the overall level set field or the stiffness of Eq.\ \eqref{ls adv}. For flows involving moderate deformations, HOUC is usually sufficient and most efficient. For more complex flows, WENO or semi-Lagragian/jet schemes combined with grid refinement might be pursed. In the present study, we use either HOUC5 or WENO5 (5 denotes fifth-order accuracy) to evaluate $\nabla \phi$.

\sloppy
For the temporal discretization of Eq.\ \eqref{ls adv}, we use a three-stage total-variation-diminishing (TVD) third-order Runge-Kutta scheme \cite{Shu_JCP_1988}. Denoting $f(\phi)=-{\bm u} \cdot \nabla \phi$, it updates $\phi$ from time level $n$ to $n+1$ in three sub-steps
\begin{equation}
  \begin{cases}
    & \phi^1 = \phi^n + \Delta t \cdot f(\phi^n)  \\
    & \phi^2 = \frac{3}{4} \phi^n + \frac{1}{4} \phi^1 + \frac{1}{4} \Delta t \cdot f(\phi^1) \\
    & \phi^{n+1} = \frac{1}{3} \phi^n + \frac{2}{3} \phi^2 + \frac{2}{3} \Delta t \cdot f(\phi^2). \\
  \end{cases}
  \label{SSP-RK3}
\end{equation}

\Ge{Finally, we note that Eq.\ \eqref{ls adv} does not need to be solved in the entire computational domain, as only the near-zero values are used to identify the interface and compute its curvature. This motivated the so-called narrow band approach \cite{Adalsteinsson_JCP_1995, Peng_JCP_1999}, which localizes the level set to the interface using index arrays. Combined with optimal data structures \cite{Nielsen_JSC_2006,Brun_JCP_2012}, fast computation and low memory footprint may be achieved at the same time. In our implementation, we store all the level set values while only update those in a narrow band, \ie solving $\phi_t+c(\phi){\bm u}\cdot \nabla \phi=0$ with the cut-off function given as}
\begin{equation}
  c(\phi) =
  \begin{cases}
     1 & \textrm{if} \quad  |\phi| < \gamma \\
     0 & \textrm{otherwise}, \\
  \end{cases}
  \label{NB cut-off}
\end{equation}
\Ge{where $\gamma= 6\Delta x$ as additional distance information is required to model droplet interactions (Sec.\ \ref{sec: idrop}). This is equivalent to \cite{Peng_JCP_1999} with a simplified $c(\phi)$.}

\paragraph {Zalesak's disk}

\begin{figure}[t!]
\centering
  \subfigure[$50\times50$]{\includegraphics[width=.3\columnwidth]{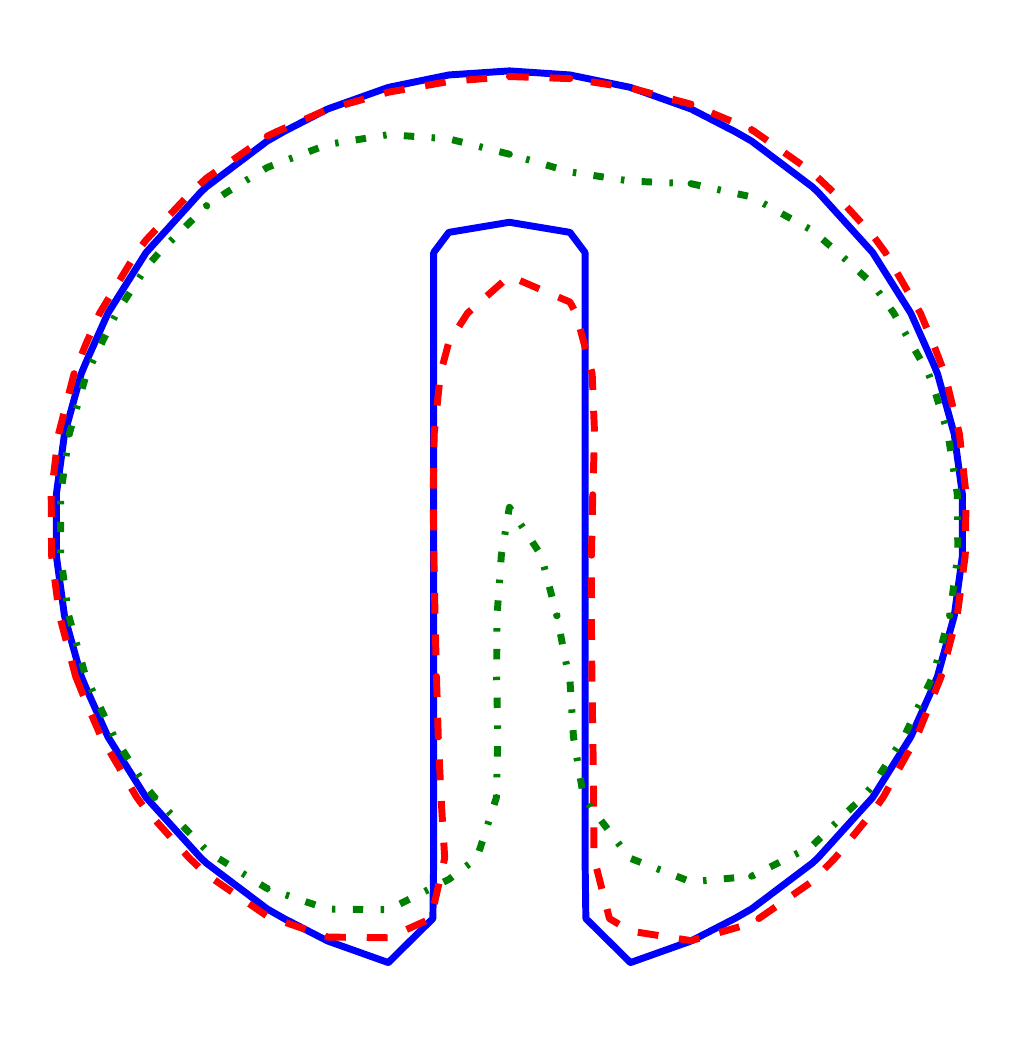}}
  \subfigure[$100\times100$]{\includegraphics[width=.3\columnwidth]{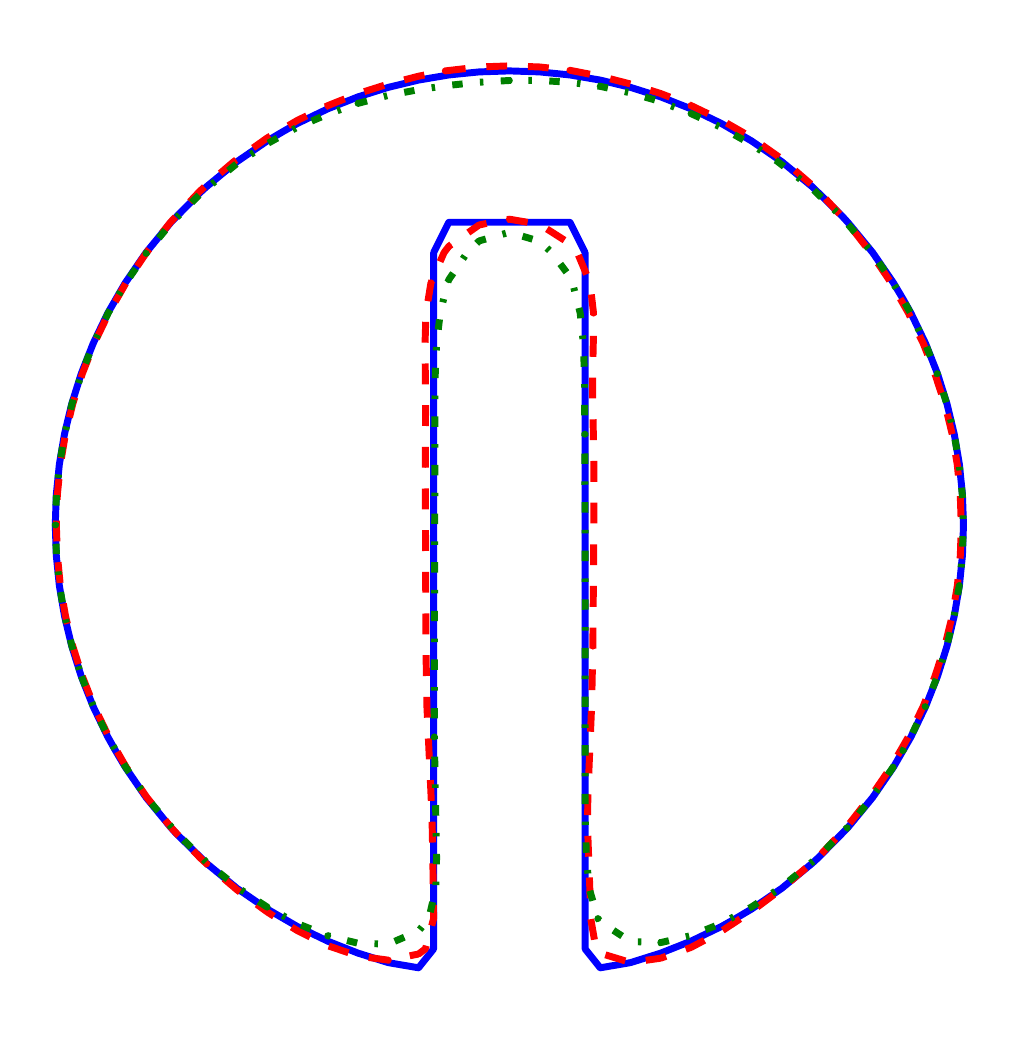}}
  \subfigure[$200\times200$]{\includegraphics[width=.3\columnwidth]{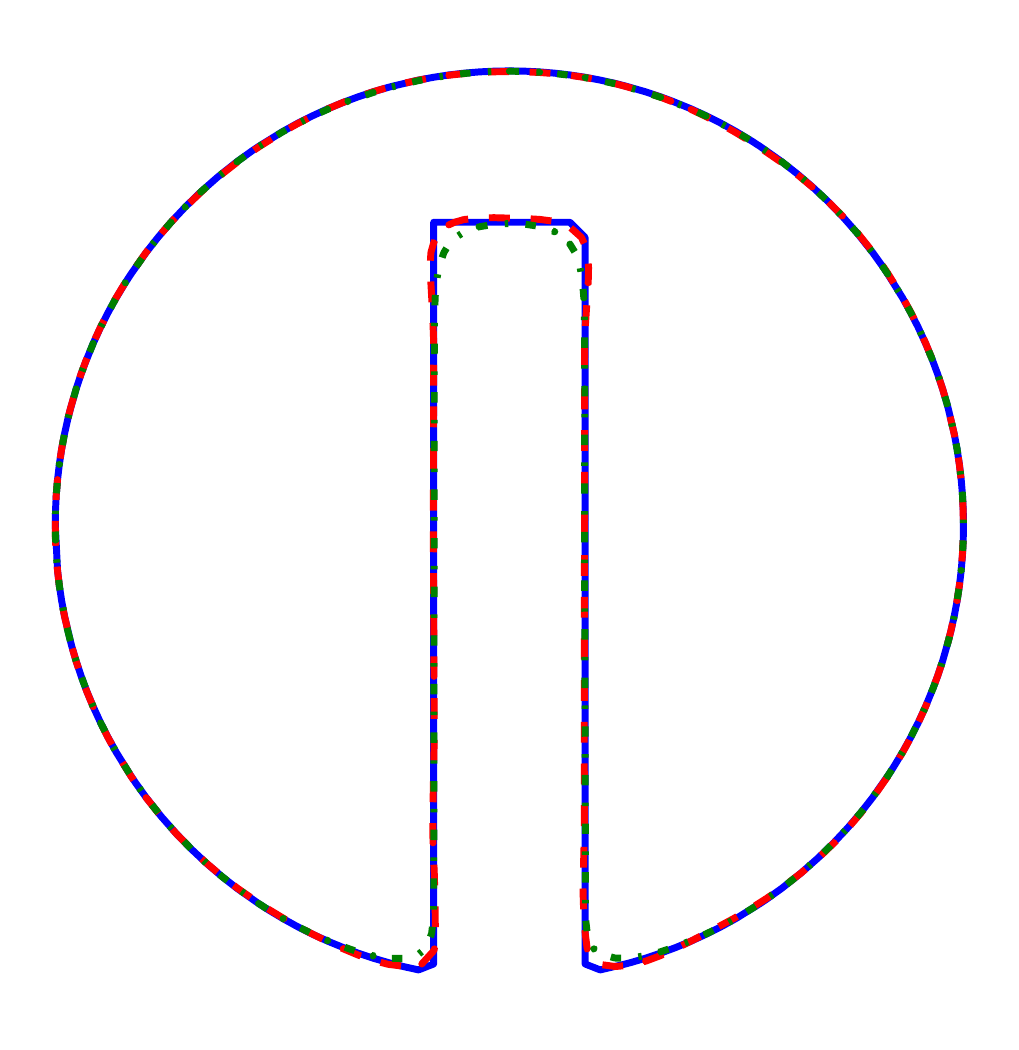}}
  \caption{Comparison of the initial interface and its shape after one full rotation for different mesh resolutions. Solid lines depict the initial interface. Two different schemes have been used to evaluate the gradients, namely HOUC5 (dashed lines) and WENO5 (dash-dotted line).}
  \label{fig:zalesak}
\end{figure}

The Zalesak's disk \cite{Zalesak_JCP_1979}, \textit{i.e.} a slotted disc undergoing solid body rotation, is a standard benchmark to validate level set solvers. The difficulty of this test lies in the transport of the sharp corners and the thin slot, especially in under-resolved cases. The initial shape should not deform under solid body rotation. Hence, by comparing the initial level set field and that after one full rotation one can characterise the degree of accuracy of a numerical solver. Here, the parameters are chosen so that a disk of radius $0.15$, slot width of $0.05$ is centered at $(x,y)=(0,0.25)$ of a $[-0.5,0.5]\times[-0.5,0.5]$ box. The constant velocity field is given as
\begin{equation}
  \begin{aligned}
  u=-2\pi y, \quad v=2\pi x.
  \end{aligned}
  \label{rot}
\end{equation}
Three different mesh resolutions have been considered, namely $50 \times 50$, $100 \times 100$ and $200 \times 200$. Fig.\ \ref{fig:zalesak} depicts the shape of the interface after one full rotation of the disk, \Ge{solving Eq.\ \eqref{ls adv} only}. Along with the results of the HOUC5 scheme (red dashed line), the shape of the interface obtained using the WENO5 scheme (green dash-dotted line) is also reported in this figure. Both schemes yield good results on fine grids, but HOUC5 clearly outperforms WENO5 on the coarsest mesh considered here. 

\subsection{Reinitialization}
\label{subsec: reinit}

\Ge{Although the level set function is initialized to be a signed-distance, it may lose this property as time evolves, causing numerical issues particularly in the evaluation of the normal and the curvature \cite{Sussman_JCP_1994}. In order to circumvent these problems, an additional treatment is required to constantly reshape $\phi$ into a distance function, \ie $|\nabla \phi| = 1$. This can be done either with a direct, fast marching method (FMM) \cite{Sethian_levelset}, or by converting it into a time-dependent Hamilton-Jacobi equation \cite{Sussman_JCP_1994} }
\begin{equation}
    \frac{\partial \phi}{\partial \tau} + S(\phi_0)(\abs{\nabla \phi} - 1) = 0,
    \label{hamilton-jacobi}
\end{equation}
where $\tau$ is a pseudo-time, and $S(\phi_0)$ is a mollified sign function of the original level set, usually defined as
\begin{equation}
    S(\phi_0) =
    \begin{cases}
     -1 & \textrm{if} \quad  \phi_0 < -\Delta x \\
      1 & \textrm{if} \quad  \phi_0 >  \Delta x \\
     \frac{\phi_0}{\sqrt{\phi_0^2 + \Delta x^2}} & \textrm{otherwise.} \\
  \end{cases}
\end{equation}

\Ge{Comparing with FMM, the second approach allows the use of higher order schemes (\eg WENO5) and is easy to parallelize; hence, it has been a much more popular choice. However, as pointed out by Russo and Smereka \cite{Russo_JCP_2000}, using regular upwinding schemes for $\nabla \phi$ near the interface does not preserve the original location of the zero level set. This can lead to mass loss, especially if the level set is far from a distance function and Eq.\ \eqref{hamilton-jacobi} needs to be evolved for long time. A simple solution is to introduce a ``subcell fix'' \cite{Russo_JCP_2000}, which pins the interface in the reinitialization by modifying the stencil. Beautifully as it works in redistancing the level set, this method is however only second order accurate and thus not well-suited for evaluating curvature. Its fourth order extension \cite{duChene_JSC_2008} suffers from stability issues and may require a very small pseudo-time step \cite{Min_JCP_2007}. Based on these observations, in this paper we solve Eq.\ \eqref{hamilton-jacobi} using the classical WENO5 \cite{Liu_JCP_1994} and the same SSP-RK3 \cite{Shu_JCP_1988}. The reinitialization is not performed at every physical time step, but depends on the advection velocity. In our applications, it typically requires one to two iterations of Eq.\ \eqref{hamilton-jacobi} per ten to a hundred time steps.}

\paragraph {Distorted elliptic field}

\begin{figure}[t!]
\centering
  \includegraphics[width=\columnwidth]{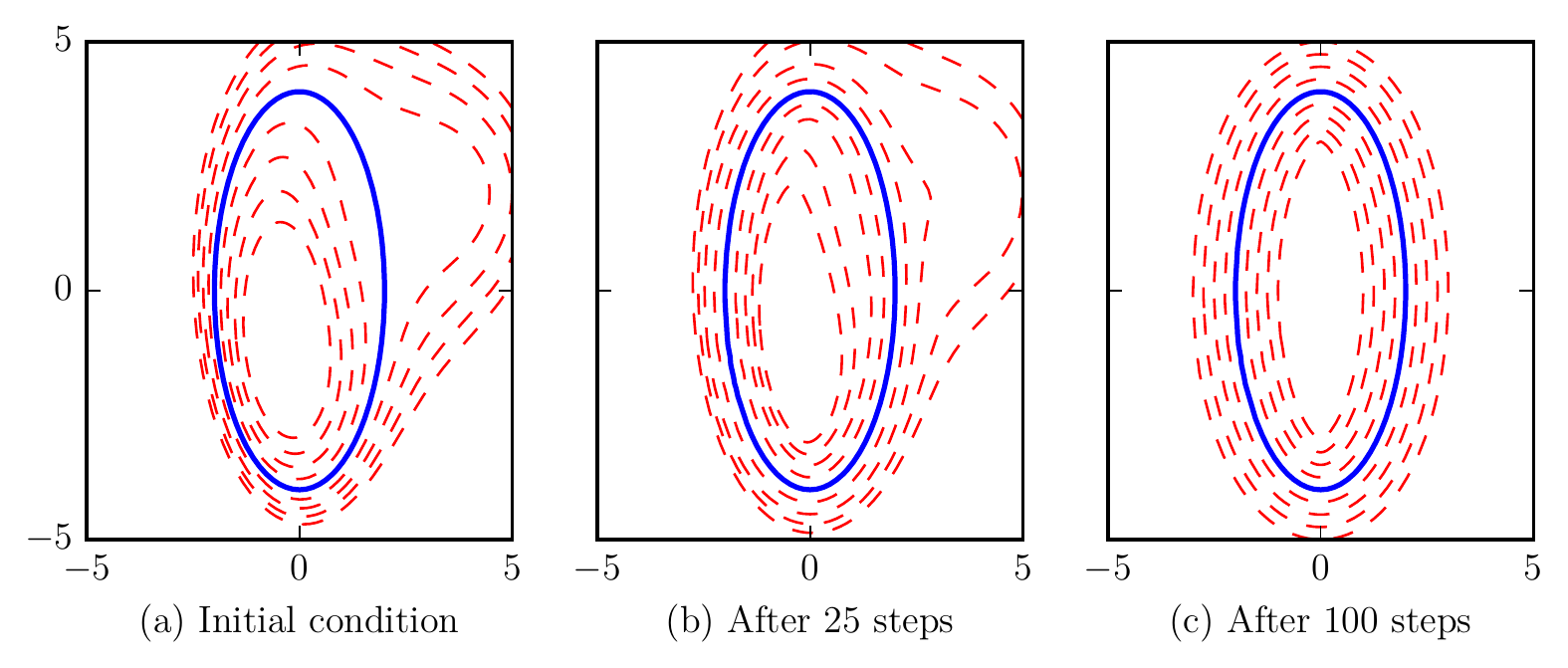}
  \caption{Illustration of the reinitialization procedure. The shape of the ellipsoid is depicted as the thick solid line. The dashed lines then depict iso-contours of $\phi(x,y)$ ranging from $-1$ to $1$ by increments of $0.25$.}
  \label{fig: redist}
\end{figure}

In order to illustrate the redistancing procedure, a test case similar to the one in \cite{Russo_JCP_2000} is considered. Define the initial level set as
\begin{equation}
  \phi(x,y,0) = f(x,y) \left( \sqrt{\left( \frac{x^2}{4} + \frac{y^2}{16} \right)} - 1 \right),
  \notag
\end{equation}
with $f(x,y)$ a distortion function that leaves only the location of the interface (an ellipse) unchanged. The initial condition is displayed in Fig.\ \ref{fig: redist}(a), where the shape of the ellipse is depicted as the thick blue line; the red dashed lines depict iso-contours of $\phi$ ranging from -1 to 1. Clearly, this initial condition is far from being equidistant. However, as $\phi(x,y,\tau)$ is evolved under Eq.\ \eqref{hamilton-jacobi}, it eventually converges towards a signed-distance function as seen in Fig.\ \ref{fig: redist}(b) and (c). 

\section{Interface-correction level set (ICLS) method}
\label{subsec: ICLS}

It is known that classical level set methods lead to mass loss when applied to multiphase flows, partially because there is no underlying mass conservation in the level set formalism, partially because of the reinitialization procedure. \Ge{Such mass loss can sometimes be reduced or even removed by using the various approaches listed in Sec.\ \ref{intro}, \eg the CLSVOF method \cite{Sussman_JCP_2000} or the hybrid particle level set method \cite{Enright_JCP_2002}.} However, doing so often makes the level set schemes complicated to implement and less efficient. To maintain the simplicity of the original level set method, we propose an alternative approach to conserve mass by performing small corrections near the interface. Because such corrections are done by directly solving a PDE (same as Eq.\ \eqref{ls adv}), the proposed method is straightforward to implement in both 2D and 3D. Meanwhile, because the correction does not need to be performed at every time step, the additional cost is also negligible. Below, we first present the derivation of the correction-velocity, then we demonstrate the mass conservation with an example.

\begin{figure}[t!]
\centering
  \includegraphics[width=.5\columnwidth]{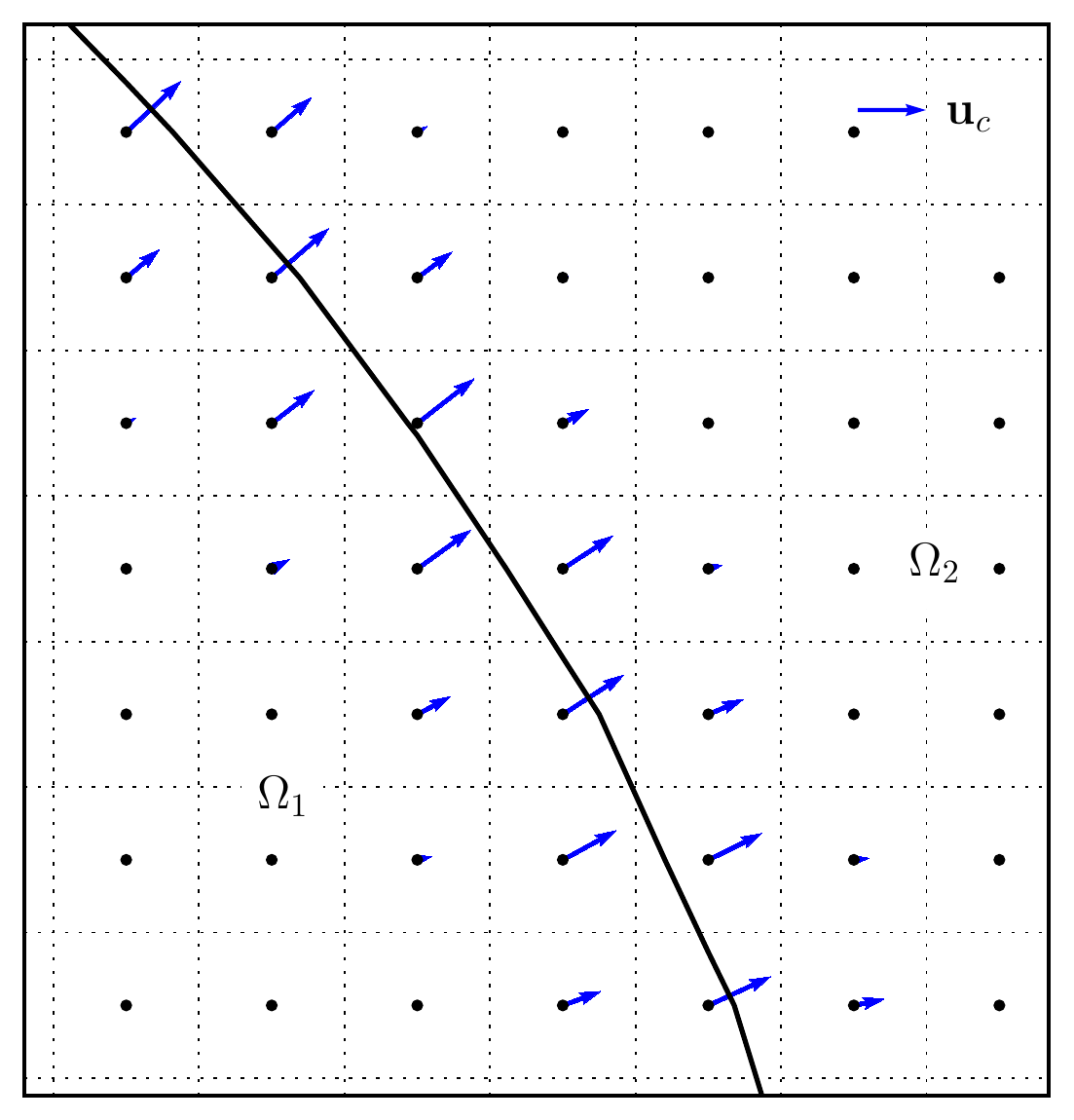}
   \caption{2D illustration of the mass correction. The solid line represents the interface. The arrows indicate the normal correction-velocity located at cell centers of the grid.}
   \label{fig:mc sketch}
\end{figure}

Let $\Gamma$ divide a domain into two disjoint subsets $\Omega_1$ (\eg a droplet) and $\Omega_2$ (\eg the ambient fluid), and $V$ denote the volume of $\Omega_1$ (Fig.\ \ref{fig:mc sketch}). Without loss of generality, we let $\phi < 0 $ in $\Omega_1$, and $\phi > 0 $ in $\Omega_2$. The rate of change of $V$ can be written as the integral of a normal velocity ${\bm u}_c$ defined on $\Gamma$ \cite{Salac_CPC_2016}, \ie
\begin{equation}
    \int_\Gamma {\bm n} \cdot {\bm u}_c \,d\Gamma = \frac{\delta V}{\delta t},
  \label{vol change}
\end{equation}
where ${\bm n}$ is the outward-pointing normal from the interface $\Gamma$. If $-\delta V/\delta t$ corresponds to the mass loss over an arbitrary period of time (it does not have to be the time step of the level set advection), then ${\bm u}_c$ can be thought as a surface velocity that corrects the volume by an amount $\delta V/\delta t$, hence compensating the mass loss. In other words, if ${\bm u}_c$ is known, then the following PDE can be solved,
\begin{equation}
  \frac{\partial \phi}{\partial t} + {\bm u}_c \cdot \nabla \phi = 0,
  \label{inflation}
\end{equation}
after which the mass loss accumulated over $\delta t$ is removed.

To obtain such a surface correction-velocity ${\bm u}_c$, we introduce a speed function $f_s$, an auxiliary pressure $p_c$, and express the rate of change of ${\bm u}_c$ as
\begin{equation}
    \frac{d{\bm u}_c}{dt} = - f_s \nabla p_c.
  \label{p2u}
\end{equation}
Here, $p_c$ can be imagined as a non-dimensional correction-pressure in $\Omega_1$. If $f_s=1$, the physical interpretation of Eq.\ \eqref{p2u} is analogous to the inflation of a balloon by $\delta V$ under pressure $p_c$ over time $\Delta t$. It is more evident rewriting ${\bm u}_c$ in the form of the impulse-momentum theorem (per unit ``mass" of the interface)
\begin{equation}
    {\bm u}_c = -\int_{0}^{\Delta t} \nabla p_c \,dt,
  \label{impulse}
\end{equation}
in which the correction-velocity is zero at $t=0$, and we require a unit speed function. In general, substituting Eq.\ \eqref{impulse} into Eq.\ \eqref{vol change} results in
\begin{equation}
    \int_0^{\Delta t}dt \int_\Gamma {\bm n} \cdot (-f_s \nabla p_c) \,d\Gamma = \frac{\delta V}{\delta t}.
  \label{ICLS 1}
\end{equation}

In order for $\nabla p_c$ to be compatible with ${\bm u}_c$, $p_c$ has to be differentiated at the interface. \Ge{Using a 1D regularized Heaviside function of $\phi$, such as}
\begin{equation}
    H_\epsilon(\phi)=
    \begin{cases}
        1 \quad \quad & \textrm{if} \quad \phi > \epsilon \\
        \tfrac{1}{2}\big[1+\tfrac{\phi}{\epsilon}+\tfrac{1}{\pi} \sin(\tfrac{\pi \phi}{\epsilon})\big] & 
        \textrm{if} \quad |\phi| \leqslant \epsilon  \\
        0 \quad & \textrm{otherwise}, \\
    \end{cases}
    \label{regularized heav}
\end{equation}
with $\epsilon=1.5\Delta x$ the half smoothing width, the correction-pressure and its gradient in Eq.\ \eqref{ICLS 1} can be conveniently written as
\begin{equation}
    p_c = \big(1-H_\epsilon(\phi)\big)p_0,
  \label{ipressure}
\end{equation}
and
\begin{equation}
    \int_\Gamma \nabla p_c = - \int_\Gamma \delta_\epsilon (\phi) \nabla \phi p_0,
  \label{ipressure grad}
\end{equation}
where $\delta_\epsilon (\phi)$ is the derivative of $H_\epsilon(\phi)$, and $p_0$ is a constant. \Ge{Note that $\bm{n} \cdot \nabla{\phi} = |\nabla \phi|$, we can denote $\int_\Gamma f_s \delta_\epsilon (\phi)|\nabla \phi| \,d\Gamma = A_f$ and express the constant pressure algebraically}
\begin{equation}
    p_0 = \frac{\delta V}{\delta t} \frac{1}{A_f \Delta t},
  \label{ipressure solution}
\end{equation}
\Ge{by substituting Eq.\ \eqref{ipressure grad} into \eqref{ICLS 1}, and approximating the time integration to first order, \ie $\int_{0}^{\Delta t} A_f \,dt = A_f \Delta t$. Finally, Eqs.\ \eqref{p2u} \eqref{ipressure grad} and \eqref{ipressure solution} can be combined to give}
\begin{equation}
    {\bm u}_c(\phi) = \frac{\delta V}{\delta t} \frac{f_s \delta_\epsilon(\phi)}{A_f} \nabla \phi,
  \label{correction vel sol}
\end{equation}
or
\begin{equation}
    {\bm u}_c(\phi) = \frac{\delta V}{\delta t} \frac{f_s}{A_f} \nabla H_\epsilon(\phi).
  \label{correction vel sol 2}
\end{equation}

\Ge{Once ${\bm u}_c$ is found, Eq.\ \eqref{inflation} can be solved for one time step to correct the mass loss.} Here, we have required a bounded support for ${\bm u}_c$, \ie ${\bm u}_c = {\bm 0}$ for $|\phi| \geqslant \epsilon$ (see Fig.\ \ref{fig:mc sketch}). There are two benefits of spreading the surface velocity. First, it allows an easy handling of the interface location, as ${\bm u}_c$ only depends on a 1D Dirac delta function of the level set. The choice of $\delta_\epsilon(\phi)$ can also be different from the trigonometric form implied from Eq.\ \eqref{regularized heav}; 
\Ge{however, we prove in \ref{appendix-dis_err} that the discretization error of $\int_\Gamma {\bm n}\cdot{\bm u}_c d\Gamma$ is always zero, independent of $\delta_\epsilon(\phi)$.}
 The important point here is we spread the \textit{correction-velocity} rather than the \textit{interface}. The interface remains sharp, as it is implicitly represented by the level set function. The second benefit of spreading ${\bm u}_c$ is that it greatly reduces the risk of numerical instability. As ${\bm u}_c$ is supported on a $2\epsilon$ band around the interface, the maximal nodal value of ${\bm u}_c$ scales with $1/\epsilon$. In our tests, we have never found its non-dimensional value to exceed 1. Therefore, the CFL conditions imposed by Eq.\ \eqref{inflation} is satisfied as long as we use the same temporal scheme (\eg RK3) for solving Eq.\ \eqref{ls adv} and Eq.\ \eqref{inflation}.
\Ge{Lastly, we remind the reader that our correction-velocity differs conceptually from the extension-velocity proposed for solving Stefan problems \cite{Chen_JCP_1997,Adalsteinsson_JCP_1999}. The extension-velocity by design will keep the level set a distance function; while the design principle here is to preserve the global mass. This distinction is clear comparing the construction procedures of the two velocities.}

\Ge{A final question is the choice of the speed function $f_s$, acting as a pre-factor for ${\bm u}_c$ in Eq.\ \eqref{correction vel sol} or \eqref{correction vel sol 2}. 
To the best of the authors knowledge, there is no simple, universally-valid criteria for such corrections. Two possible ways are}
\begin{equation}
    f_s \equiv
    \begin{cases}
        1            & \textrm{uniform speed} \\
        \kappa(\phi) & \textrm{curvature-dependent speed}. \\
    \end{cases}
    \label{speed func}
\end{equation}
The uniform speed will obviously result in a fixed strength $\delta V/\delta t /A_f$ for the velocity distribution. In the case of a static spherical droplet, this is the ideal choice for $f_s$, since the droplet should remain a sphere. In more general cases, when a fluid interface is subject to deformations or topological changes, a curvature-dependent speed may be more appropriate. \Ge{This is based on the assumption that local structures of higher curvature or regions where the flow characteristics merge tend to be under-resolved \cite{Enright_JCP_2002}; hence, they are more prone to mass losses. Indeed, a linear curvature weight has been adopted by many and demonstrated to produce accurate results in different contexts \cite{Luo_JCP_2015,Aanjaneya_JCP_2013}.} Furthermore, $\kappa/A_f$ reduces to $1/A_f$ when the curvature is uniform. Therefore, we can rewrite Eq.\ \eqref{correction vel sol 2} using a curvature-dependent speed
\begin{equation}
    {\bm u}_c(\phi) = \frac{\delta V}{\delta t} \frac{\kappa(\phi)}{A_f} \nabla H_\epsilon(\phi).
  \label{correction vel sol *}
\end{equation}
Clearly, this correction-velocity is larger in highly curved parts, and smaller in flatter parts. It thus includes ``local" information while maintaining ``global" mass conservation. Standard central-difference discretization applies, where the components of ${\bm u}_c$ can be obtained at either the cell faces or cell centers. The computation of $\kappa(\phi)$ is crucial and will be presented in the next section. We stress that such a curvature-dependence is not unique. In principle, one can choose different weight-functions, \Ge{and validate the choice based on the specific applications}. Practically, the difference is expected to be negligible since the mass loss remains small (typically around $10^{-5}$) at each correction step.

\Ge{After correcting the level set on a $2\epsilon$ band around the interface, a reinitialization step is required to redistance the values within the entire narrow band ($2\gamma$).} The two procedures can be readily combined, since it is not necessary to perform mass correction at every time step. Also, because the formalism is cast in a level set frame, generalization from 2D to 3D is trivial. \Ge{Comparing with other mass-preserving methods, the additional computational cost of ICLS is small. This is due to the simple algebraic expression of ${\bm u}_c$ (Eq.\ \eqref{correction vel sol *}), and only one solve of Eq.\ \eqref{inflation} is required}; whereas a typical VOF-coupling method involves solving another set of transport equations \cite{Sussman_JCP_2000}, or reconstructing the interface by an iterative procedure \cite{Luo_JCP_2015}.

In summary, the ICLS method proceeds by performing the following steps:

\begin{enumerate}
    \item Advect $\phi^{n}$ from time $t^{n}$ to $t^{n+1}$ with Eq.\ \eqref{ls adv}, using the flow velocity ${\bm u}^{n}$.
    
    \item If reinitialization will be executed (otherwise, go to step 3):
      \begin{enumerate}
        \item Perform mass correction with Eq.\ \eqref{inflation}, using ${\bm u}_c$ from Eq.\ \eqref{correction vel sol *}.
        \item Reinitialize $\phi^{n+1}$ with Eq.\ \eqref{hamilton-jacobi}.
      \end{enumerate}
    
    \item Exit the level set solver.
\end{enumerate}

\paragraph {Deforming circle}

\begin{figure}[t]
\centering
  \includegraphics[width=.9\columnwidth]{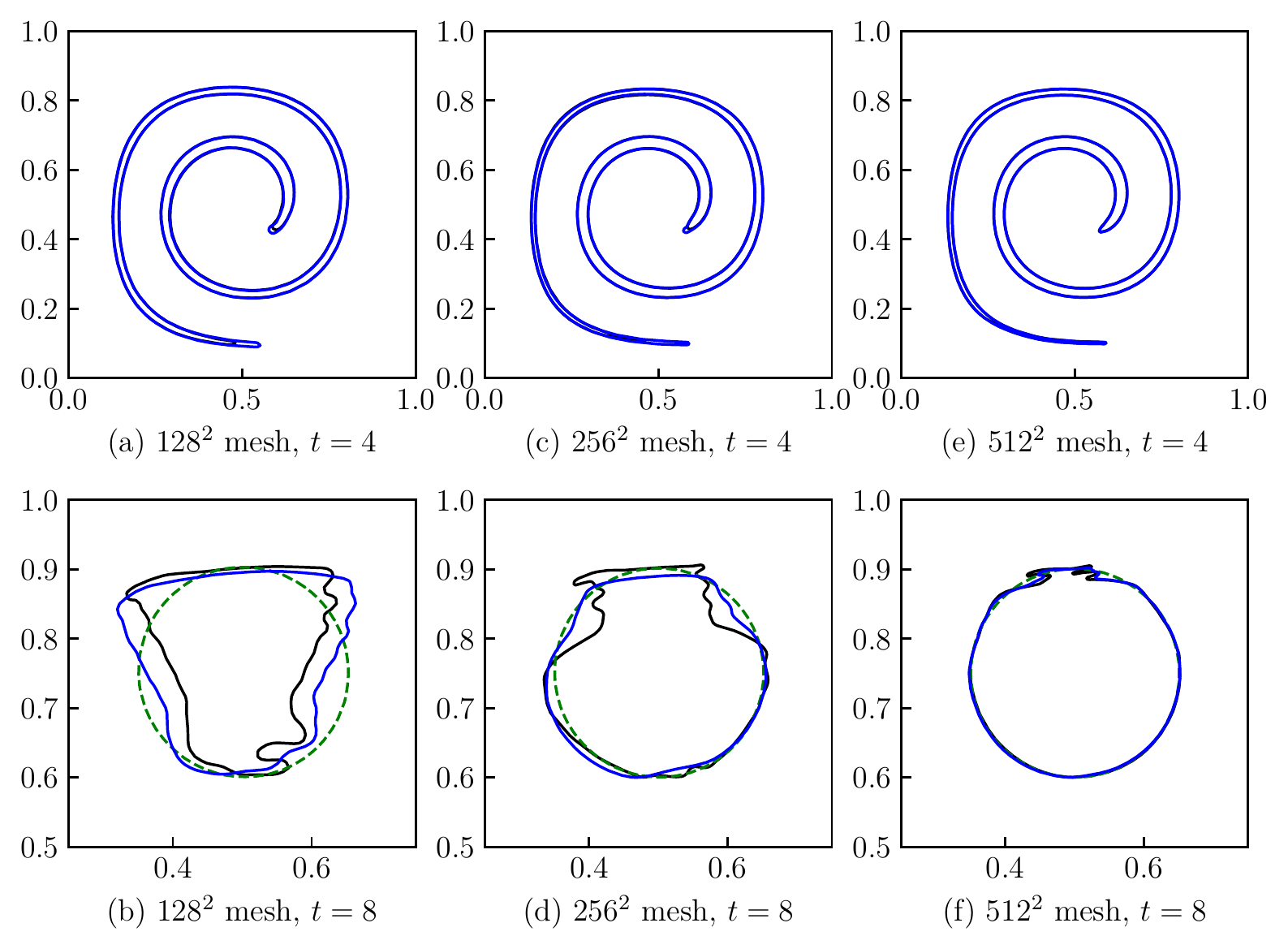}
   \caption{Interface at $t=4$ and $t=8$ for different meshes. The solid black lines indicate simulations without mass correction, the solid blue lines indicate simulations with the current mass correction method, the green dashed lines in (b)(d)(f) indicate the original circle. (For interpretation of the references to color in this figure legend, the reader is referred to the web version of this article.)}
   \label{fig:serpentine}
\end{figure}

\begin{figure}[t]
 \begin{center}
 \includegraphics[width=\columnwidth]{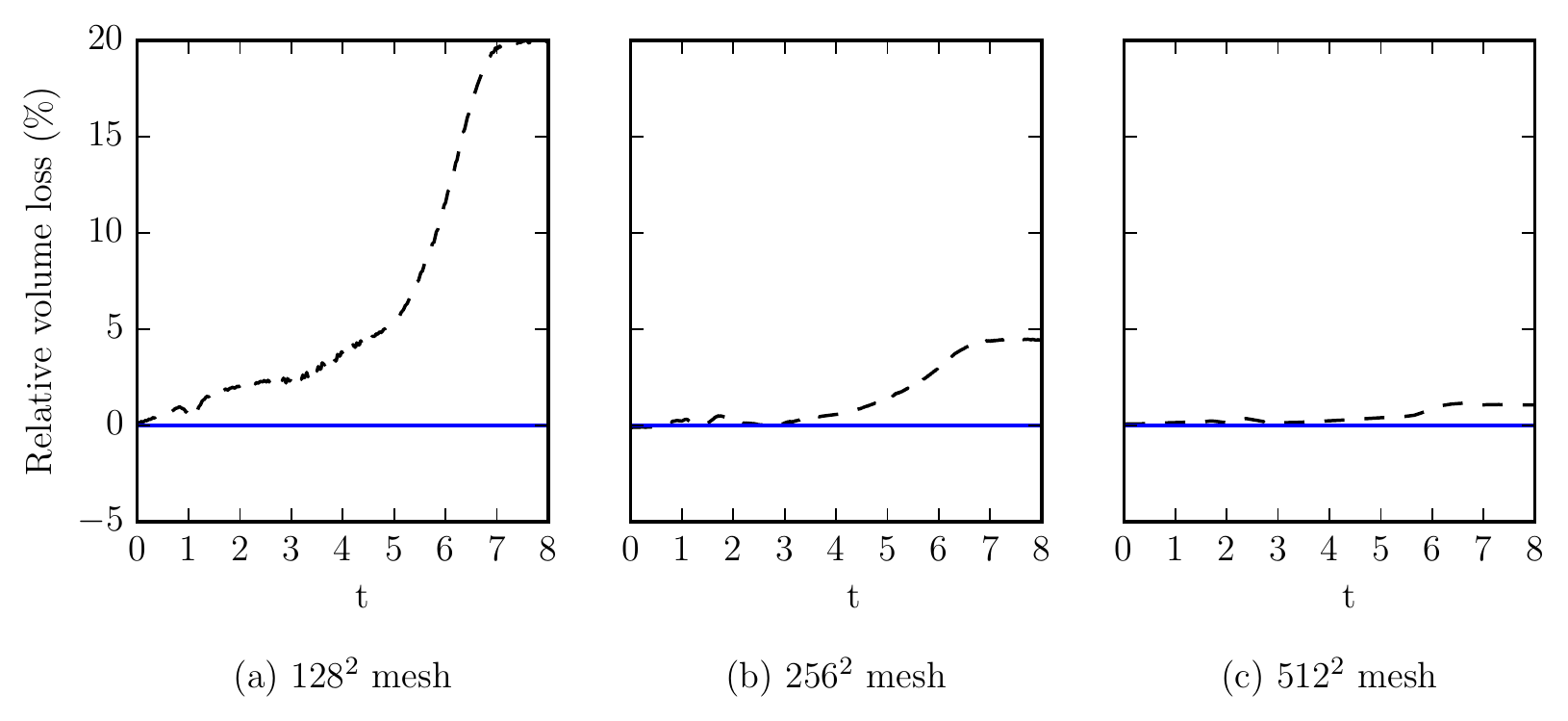}
 \end{center}
 \caption{Relative volume loss for three different meshes. Dashed lines indicate simulations without mass correction; solid lines indicate simulations with mass correction.}
 \label{fig: massloss}
\end{figure}

\Ge{To assess the performance of ICLS on mass conservation, we test the standard benchmark of a circle deformed by a single vortex. Here, the circle of radius $0.15$ is initially centered at $(x,y)=(0.5, 0.75)$ of a $\left[ 0, 1 \right] \times \left[ 0, 1 \right]$ box. The velocity is imposed directly and can be obtained from the stream function}
\begin{equation}
  \psi(x,y,t) = \frac{1}{\pi} \sin^2(\pi x) \sin^2(\pi y) \cos \left( \frac{\pi t}{T} \right),
  \notag
\end{equation}
where $T$ is traditionally set to 8. \Ge{Under this flow, the circle will be stretched to maximum at $t = T/2$ and rewound to its initial condition at $t=T$. Although formulated simply, accurately transporting the interface without mass loss is a difficult task.}

\Ge{We perform this test on three different meshes using the complete level set solver: HOUC5 is used for the level set advection, WENO5 is used for reinitialization every 5 to 20 time steps, the mass correction is performed every 5 to 10 time steps; and the time step is chosen such that $\Delta t/\Delta x=0.32$. Fig.\ \ref{fig:serpentine} shows the shapes of the filament/circle at $t=4$ and $t=8$ at various resolutions. From the upper panel, it is clearly seen that the filament has a longer tail and head due to mass correction; as we increase the resolution, the difference becomes smaller. The lower panel of Fig.\ \ref{fig:serpentine} depicts the final shapes, ideally the initial circle if the motion is totally passive. Some artifacts are visible due to the fact that the filament is always under-resolved at the maximum stretching and the level set will automatically merge the characteristics to yield an entropy solution \cite{Sethian_levelset}. We note that the final outcome can be tuned by modifying the frequency of the reinitialization/mass correction, a trade-off between the appearance and the mass loss. However, the objective here is to demonstrate the mass conservation enforced by ICLS, which is clearly illustrated in Fig.\ \ref{fig: massloss}. For passive transport involving large deformations, we recommend particle-based methods \cite{Enright_JCP_2002}. Examples of droplets/bubbles in physical conditions using ICLS will be shown in the validations (Sec.\ \ref{sec: validations}) and applications (Sec. \ref{sec: idrop}) below.} 


\section{Curvature computation}
\label{subsec: curv}

\Ge{Curvature computation is crucial to interfacial flows in the presence of surface tension, as inaccurate curvature can result in unphysical spurious currents \cite{Herrmann_JCP_2008, Desjardins_JCP_2008}, and even more so in our case when we apply curvature-dependent interface corrections. 
In this section, we first briefly describe the calculation of cell-center curvatures; \textit{i.e.}, the curvature evaluated at the same nodal position as the level set function. Then, we introduce a geometric approach for the estimation of interface curvatures corresponding to the zero level set. The second step is specially tailored to the ghost fluid method that will be presented in Sec.\ \ref{subsec: gfm}.}

\subsection{Cell-center curvature}
\label{subsec: cc curv}

From Eq.\ \eqref{curv}, the curvature $\kappa$ can be evaluated as

\begin{equation}
  \begin{aligned}
    \kappa =- \frac{\phi_{yy}\phi_x^2 + \phi_{xx}\phi_y^2 - 2\phi_x\phi_y\phi_{xy}}{{(\phi_x^2+\phi_y^2) }^{3/2}}
  \end{aligned}
  \label{curv 2d}
\end{equation}
and as
\begin{equation}
  \kappa_M = -
  \begin{aligned}
    \frac{\bigg\{ 
          \begin{aligned}
          (\phi_{yy}+\phi_{zz})\phi_x^2 + (\phi_{xx}+\phi_{zz})\phi_y^2  + (\phi_{xx}+\phi_{yy})\phi_z^2 \\
          - 2\phi_x\phi_y\phi_{xy} - 2\phi_x\phi_z\phi_{xz} - 2\phi_y\phi_z\phi_{yz}
          \end{aligned}
          \bigg\}
          } {{(\phi_x^2+\phi_y^2+\phi_z^2) }^{3/2}}
  \end{aligned}
  \label{curv 3d}
\end{equation}
in 2D and 3D Cartesian coordinates, respectively, where the subscript $M$ denotes the mean curvature \cite{Sethian_levelset}. The curvature can be determined from these expressions using simple central finite-differences. It has to be noted, however, that such evaluation of $\kappa$ involves second derivatives of the level set field $\phi({\bm x})$. As a consequence, if the \Ge{calculation} of $\phi$ is only second-order accurate, the resulting $\kappa$ will be of order zero. To nonetheless retain a grid converging $\kappa$, one can use the compact least-squares scheme proposed by Marchandise \etal \cite{Marchandise_JCP_2007}. Their approach provides a second-order, grid converging evaluation of the cell-center curvature. It moreover smears out undesired high frequency oscillations possibly introduced by the velocity field. A similar procedure has also been adopted in other works \cite{Desjardins_JCP_2008, Luo_JCP_2015}.

The principle of the least squares approach is to solve an over-determined linear system, $\bm {A x} = {\bm b}$, where ${\bm A}$ is a matrix built from the local coordinates, ${\bm x}$ is a unknown array containing the reconstructed level set values and its spatial derivatives, and ${\bm b}$ is the original level set field. The detailed descriptions can be found in \cite{Marchandise_JCP_2007}. Here, we only note that the level set function remains unmodified after this step. From a practical point of view, provided the mesh considered is uniform in all directions, the pseudo-inverse of the matrix ${\bm A}$ only needs to be evaluated once and applied close to the interface. Therefore, the computational cost of this least-squares calculation is negligible.

\subsection{Interface curvature}

The least-squares approach described in the previous section only allows one to compute the nodal curvature $\kappa$ of the level set field $\phi$. For computations using the GFM (Sec.\ \ref{subsec: gfm}), one might however require an accurate evaluation of the curvature at the exact location of the interface. Provided a grid-converging cell-center curvature, the actual curvature at the interface can be interpolated from its neighboring cells weighted by the level set \cite{Francois_JCP_2006, Luo&Hu_JCP_2015}. Here we present a slightly different but robust algorithm to estimate the interface curvature, with a straight-forward geometrical interpretation. 

\begin{figure}[t!]
 \begin{center}
 \includegraphics[width=.75\columnwidth]{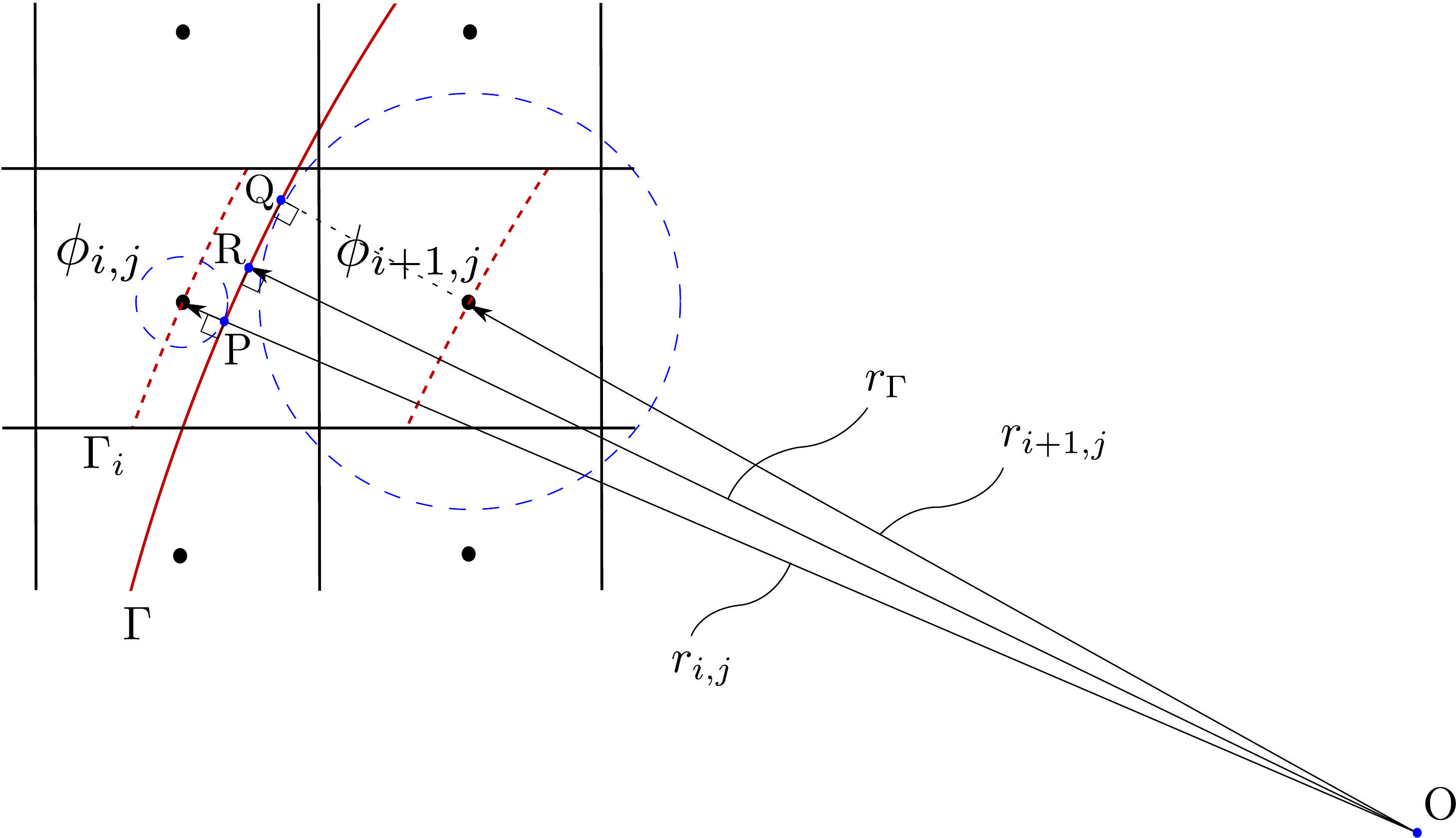}
 \end{center}
 \caption{Estimation of the interface's curvature from neighboring cells.}
 \label{fig: icurv}
\end{figure}

\paragraph{2D estimation}

Suppose the interface $\Gamma$ cuts through two adjacent cells, $(i,j)$ and $(i+1,j)$, where the cell-center curvatures $\kappa_{i,j}$ and $\kappa_{i+1,j}$ are known. In 2D, we can determine the radius of curvature at each cell directly from
\begin{equation}
    \kappa_{i,j} = -\frac{1}{r_{i,j}}, \quad\quad \kappa_{i+1,j} = -\frac{1}{r_{i+1,j}},
  \label{2d curv}
\end{equation}
as illustrated in Fig.\ \ref{fig: icurv}. 
Since the level set is defined as the signed distance to the interface, $\Gamma$ must be tangent to a circle of radius $\abs{\phi_{i,j}}$ centered at $(i,j)$, and parallel to the contour line of $\Gamma_i = \{{\bm x} |\phi=\phi_{i,j}\}$ (otherwise they will not remain equidistant). We also know $\Gamma$ lies between $(i,j)$ and $(i+1,j)$, then it must pass through $P$ (see Fig.\ \ref{fig: icurv}). Since $\Gamma$ and $\Gamma_i$ are parallel and there is only one line normal to both curves passing through P, $r_{i,j}$ and $OP$ must originate from the same point, $O$. Then we get
\begin{equation}
    \abs{OP} = r_{i,j} - s_\Gamma \phi_{i,j}.
  \label{2d curv 1}
\end{equation}
where $s_\Gamma$ is a sign function equal to $1$ if the interface wrapping the negative level set is convex, and equal to $-1$ if concave.

The same argument holds for cell $(i+1,j)$, which yields $\abs{OQ} = r_{i+1,j} - s_\Gamma \phi_{i+1,j}$. We can therefore write the radius of the interface curvature between $(i,j)$ and $(i+1,j)$ as
\begin{equation}
    r_\Gamma = \frac{\abs{OP}+\abs{OQ}}{2},
  \label{2d icurv 0}
\end{equation}
so that the interface curvature becomes
\begin{equation}
    \kappa_{\Gamma} = \displaystyle \frac{2}{ \kappa_{i,j}^{-1} + \kappa_{i+1,j}^{-1} +s_\Gamma (\phi_{i,j}+\phi_{i+1,j}) }.
  \label{2d icurv}
\end{equation}

The above derivation provides a relation between the interface curvature and that at the adjacent cell-centers in the $x$ direction. Similar results can be obtained in the $y$ direction (\eg between $\phi_{i,j}$ and $\phi_{i,j-1}$). The assumptions we have made here are 1) the cell-center curvatures are accurate and 2) the interface curvatures at $P$ and $Q$ are the same, so that $OP$ and $OQ$ are co-centered (or, $\abs{OP} \approx \abs{OQ} \approx \abs{OR}$). The second assumption is essentially a sub-cell approximation, and we expect it to be valid as long as the interface is well-resolved. One exception we have found is when two interfaces are closer than about $2\Delta x$, the local level set field will develop ``corners". In that case, the cell-center curvatures are erroneous and the underlying assumptions we require here are not fulfilled. We do not discuss that case in the present paper. However, we demonstrate in the next section that a second-order convergence is achieved when the interface is resolved.

\paragraph{3D estimation}

In three dimensions, the mean curvature of a surface can be written as
\begin{equation}
  \begin{aligned}
    \kappa_\Gamma = -(\frac{1}{r_{\Gamma1}}+\frac{1}{r_{\Gamma2}}),
  \end{aligned}
  \label{3d icurv def}
\end{equation}
where $r_{\Gamma1}$ and $r_{\Gamma2}$ are the two principal radii corresponding to the maximal and minimal planar radius of curvature. Note that we do not need to approximate the interface as a sphere since there is always a plane where the previous picture (Fig.\ \ref{fig: icurv}) holds. Under the same assumption as for the 2D case, that the interface at $P$ and $Q$ have the same principal radii (hence the same curvature), one can again relate the nodal curvatures to their nearby interface as
\begin{equation}
  \begin{aligned}
    & \kappa_{i,j,k} = -(\frac{1}{r_{\Gamma1}+s_\Gamma\phi_{i,j,k}}+\frac{1}{r_{\Gamma2}+s_\Gamma\phi_{i,j,k}}), \\
    &\kappa_{i+1,j,k} = -(\frac{1}{r_{\Gamma1}+s_\Gamma\phi_{i+1,j,k}}+\frac{1}{r_{\Gamma2}+s_\Gamma\phi_{i+1,j,k}}),
  \end{aligned}
  \label{3d curv}
\end{equation}
where $s_\Gamma$ is the same sign function defined for the 2D case. 
Comparing equations \eqref{3d icurv def} and \eqref{3d curv}, it is natural to expand Eq.\ \eqref{3d curv} into a Taylor series and to approximate the interface curvature directly as 
\begin{equation}
    \kappa_\Gamma = \frac{\epsilon_{i+1}\kappa_i - \epsilon_i\kappa_{i+1}}{\epsilon_{i+1} - \epsilon_i} +O(\epsilon_i^2,\epsilon_{i+1}^2),
  \label{3d icurv2}
\end{equation}
\noindent where
\begin{equation}
    \epsilon_i = s_\Gamma\phi_{i,j,k}.
  \label{none}
\end{equation}
Since the level set must change sign across the interface, Eq.\ \eqref{3d icurv2} is always defined and it reduces to the exact value if the cell center happens to be on the interface. Similarly, the whole procedure is repeated in the $y$ and $z$ directions.

Finally, in order to ensure a robust estimation, we perform an additional quadratic least squares approximation on the curvature field near the interface, similar to \cite{Marchandise_JCP_2007}. This procedure takes place before the 3D estimation (Eq.\ \eqref{3d icurv2}), and essentially improves the accuracy of cell-center curvatures by removing possible high-frequency noise. \Ge{We note that the second averaging is optional, and different methods can be found in literature to evaluate the cell-center curvatures \cite{duChene_JSC_2008}. In the present paper, the least squares approach mentioned in Sec.\ \ref{subsec: cc curv} is used for all the cases.}

\begin{table}[t]
    \centering
    \caption{Grid convergence of the current interface curvature calculation in both 2D and 3D.}
    \tabulinesep=1.2mm
    \begin{tabular}{ l l l l l }
        \hline
        Grid points per diameter    &16          &32          &48          &64   \\
        \hline
        $L_\infty$ \quad 2D         &1.144\e{-2} &2.904\e{-3} &1.285\e{-3} &7.227\e{-4}  \\
        $L_\infty$ \quad 3D         &1.527\e{-2} &3.888\e{-3} &1.732\e{-3} &9.753\e{-4}  \\
     \hline
 \end{tabular}
 \label{tab: icurv linf}
\end{table}

\begin{figure}[t]
\centering
  \includegraphics[width=.5\columnwidth]{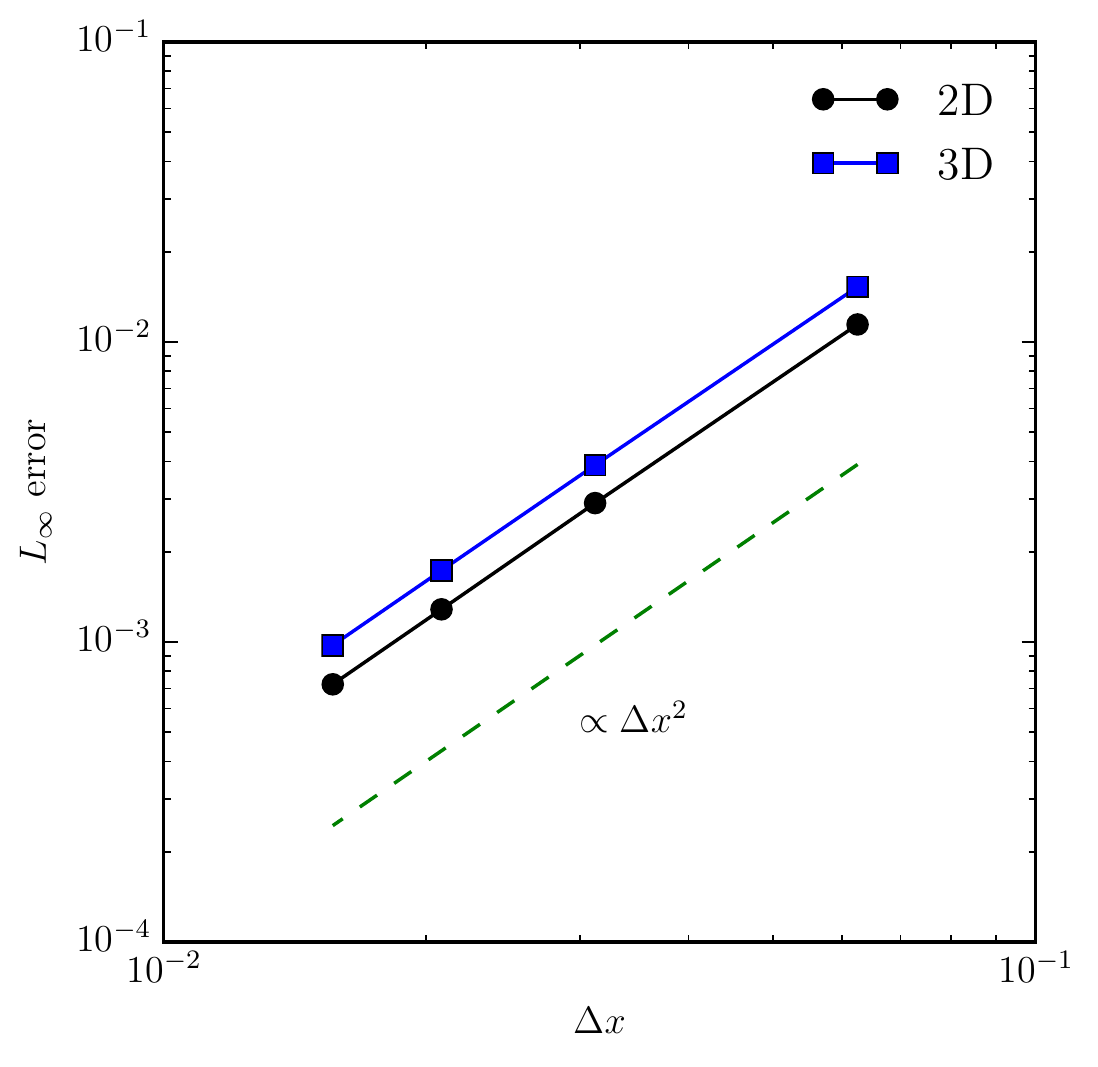}
   \caption{Second order convergence of the interface curvature computation in both 2D and 3D.}
   \label{fig: icurv norm}
\end{figure}

To assess the accuracy of our interface curvature estimation, we calculate the $L_\infty$ norm of a circle/sphere of radius $0.25$ centered in a unit square/cube. Table \ref{tab: icurv linf} summarizes the error after one step of the calculations on different resolutions, which are also plotted in Fig.\ \ref{fig: icurv norm}. Clearly, second-order convergence is achieved in both 2D and 3D cases.


\section{Solution of the Navier-Stokes equations}
\label{subsec: NS}

In this section, we outline the flow solver developed from that of Breugem \cite{Wim-Paul_JCP_2012} for particle-laden flows. After advancing the level set from $\phi^{n}$ to $\phi^{n+1}$, the density and viscosity fields are updated by
\begin{subequations}
 \begin{equation}
  \rho^{n+1} = \rho_1 H_s(\phi^{n+1}) + \rho_2 (1-H_s(\phi^{n+1})), 
 \end{equation}
 \begin{equation}
  \mu^{n+1} = \mu_1 H_s(\phi^{n+1}) + \mu_2 (1-H_s(\phi^{n+1})), 
  \label{visc profile}
 \end{equation}
\end{subequations}
\noindent where
\begin{equation}
    H_s(\phi)=
    \begin{cases}
        1 \quad \quad \textrm{if} \quad \phi > 0 \\
        0 \quad \quad \textrm{otherwise}, \\
    \end{cases}
\end{equation}
\noindent is a simple step function.

Next, a prediction velocity ${\bm u}^*$ is computed by defining ${\bf RU}^n$ as
\begin{equation}
  {\bf RU}^n = -\nabla \cdot ({\bm u}^n {\bm u}^n)+\frac{1}{Re}\bigg(\frac{1}{\rho^{n+1}} \nabla \cdot \big[\mu^{n+1}(\nabla {\bm u}^n+(\nabla {\bm u}^n)^T)\big]\bigg)+\frac{1}{Fr}{\bm g},
  \label{ru}
\end{equation}
\noindent which is the right-hand side of the momentum equation \eqref{NS} excluding the pressure gradient term. Integrating in time with the second-order Adams-Bashforth scheme \Ge{(AB2)} yields
\begin{equation}
    {\bm u}^* ={\bm u}^n +\Delta t \bigg(\frac{3}{2} {\bf RU}^n-\frac{1}{2} {\bf RU}^{n-1} \bigg).
  \label{ab2}
\end{equation}

To enforce a divergence-free velocity field (Eq.\ \eqref{div free}), we proceed by solving the Poisson equation for the pressure as in the standard projection method \cite{Chorin_1968}, \ie
\begin{equation}
  \nabla \cdot \bigg(\frac{1}{\rho^{n+1}} \nabla p^{n+1} \bigg) = \frac{1}{\Delta t}\nabla \cdot {\bm u}^*.
  \label{var poisson}
\end{equation}
\noindent The surface tension between two fluids is also computed during this step, using the ghost fluid method \cite{Fedkiw_JCP_1999} (Sec.\ \ref{subsec: gfm}). This allows for an accurate and sharp evaluation of the pressure jump even at large density contrasts \cite{Desjardins_JCP_2008}. Finally, the velocity at the next time level is updated as
\begin{equation}
    {\bm u}^{n+1} = {\bm u}^* -\frac{\Delta t}{\rho^{n+1}} \nabla p^{n+1}.
  \label{projection}
\end{equation}


\subsection{Fast pressure-correction method}
\label{p correction}

In the above outline, a Poisson equation for the pressure (Eq.\ \eqref{var poisson}) must be solved at each time step. This operation takes most of the computational time in the projection method, as it is usually solved iteratively. In addition, the operation count of iterative methods depends on the problem parameters (\eg density ratio) and the convergence tolerance \cite{Dodd_JCP_2014}. On the other hand, Dong and Shen \cite{Dong_JCP_2012} recently developed a velocity-correction method that transforms the variable-coefficient Poisson equation into a constant-coefficient one. The essential idea is to split the pressure gradient term in Eq.\ \eqref{var poisson} in two parts, one with constant coefficients, the other with variable coefficients, \ie
\begin{equation}
    \frac{1}{\rho^{n+1}} \nabla p^{n+1} \rightarrow \frac{1}{\rho_0} \nabla p^{n+1} + ( \frac{1}{\rho^{n+1}} - \frac{1}{\rho_0} ) \nabla \hat{p},
  \label{pressure split}
\end{equation}
where $\rho_0=min(\rho_1,\rho_2)$ and $\hat{p}$ is the approximate pressure at time level $n+1$. This splitting reduces to the exact form of Eq.\ \eqref{var poisson} within the lower-density phase, while its validity in the higher-density phase and at the interface depends on the choice of $\hat{p}$. Later, Dodd and Ferrante \cite{Dodd_JCP_2014} showed that by explicitly estimating $\hat{p}$ from two previous time levels as
\begin{equation}
    \hat{p} = 2p^{n}-p^{n-1},
  \label{pressure approx}
\end{equation}
the resulting velocity field in Eq.\ \eqref{projection} will be second-order accurate in both space and time, 
\Ge{independent of the interface advection method}.
Furthermore, 
\Ge{if the computational domain includes periodic boundaries or can be represented by certain combination of homogeneous Dirichlet/Neumann conditions \cite{Schumann_JCP_1988},}
the constant-coefficient part of Eq.\ \eqref{pressure split} can be solved directly using Gauss elimination in the Fourier space. Such a FFT-based solver can lead to a speed-up of $10-40$ times, thus the name fast pressure-correction method (FastP*). Following this approach, Eqs.\ \eqref{var poisson} and \eqref{projection} are modified as
\begin{equation}
  \nabla^2 p^{n+1} = \nabla \cdot \bigg[ \big(1-\frac{\rho_0}{\rho^{n+1}}) \nabla \hat{p} \bigg] +  \frac{\rho_0}{\Delta t}\nabla \cdot {\bm u}^*
  \label{poisson split}
\end{equation}
\noindent and
\begin{equation}
    {\bm u}^{n+1} = {\bm u}^* -\Delta t \bigg[\frac{1}{\rho_0} \nabla p^{n+1} + \big(\frac{1}{\rho^{n+1}} - \frac{1}{\rho_0}\big)\nabla \hat{p} \bigg].
  \label{projection 1}
\end{equation}


\subsection{Ghost fluid method}
\label{subsec: gfm}

As discussed before, surface tension is commonly computed using the continuum surface force (CSF) model \cite{Brackbill_JCP_1992}, in which the pressure jump across an interface is represented as a forcing term on the right-hand side of Eq.\ \eqref{NS}. Despite its simplicity, CSF introduces an unfavorable smearing in the density and pressure profiles, resulting in an artificial spreading of the interface (typically over a thickness of $3\Delta x$). An alternative approach is the so-called ghost fluid method (GFM), originally developed by Fedkiw \etal \cite{Fedkiw_JCP_1999} to capture the boundary conditions in the inviscid compressible Euler equations. Unlike CSF, GFM enables a numerical discretization of the gradient operator while preserving the discontinuity of the differentiated quantity. It was extended to viscous flows by Kang \etal \cite{Kang_JSC_2000} and has been successfully utilized in multiphase flow simulations, see \eg  \cite{Desjardins_JCP_2008, Coyajee_JCP_2009, Tanguy_2005}. 

Recall from Eq.\ \eqref{pressure jump} that the pressure jump has two components, one arising from the surface tension, the other from the viscosity difference of the two fluids. In \cite{Kang_JSC_2000}, a complete algorithm is provided to compute the two contributions, making the density, viscosity, and pressure all sharp. However, having a sharp viscosity profile requires an extra step to evaluate the divergence of the deformation tensor (see Eq.\ \eqref{ru}). That is, for cells adjacent to the interface, the second derivatives of the velocity must be evaluated using the techniques developed in \cite{Liu_JCP_2000, Kang_JSC_2000}. 
However, rewriting Eq.\ \eqref{pressure jump} as
\begin{equation}
    [p]_\Gamma = \frac{1}{Re} \bigg(\frac{\kappa}{Ca} + 2[\mu]_\Gamma {\bm n}^T \cdot \nabla {\bm u} \cdot {\bm n} \bigg),
  \label{pressure jump 2}
\end{equation}
\noindent reveals that surface tension is the dominant term when the Capillary number, $Ca=We/Re$, is small. For the applications we are interested in, \eg colloidal droplets in microfluidic channels, $Ca$ is of the order of $10^{-5}$. Therefore, in the present implementation, we regularize the viscosity profile (\ie replacing $H_s(\phi)$ in Eq.\ \eqref{visc profile} with $H_\epsilon(\phi)$ in Eq.\ \eqref{regularized heav}) and use GFM only for the pressure jump. 


\subsubsection{Spatial discretization}
\label{sec: spatial}

Eqs.\ \eqref{ru}, \eqref{poisson split}, and \eqref{projection 1} are discretized on a standard staggered grid using a second-order conservative finite volume method. It is equivalent to central differences in all three directions if the mesh is uniform. A detailed description of the discretization of the individual terms can be found in \cite{Dodd_JCP_2014}, Sec.\ 2.2.1. For brevity, we show here only the 2D evaluations of $\nabla p$ and $\nabla^2 p$ due to GFM.
\begin{figure}[t]
 \begin{center}
 \includegraphics[width=5cm]{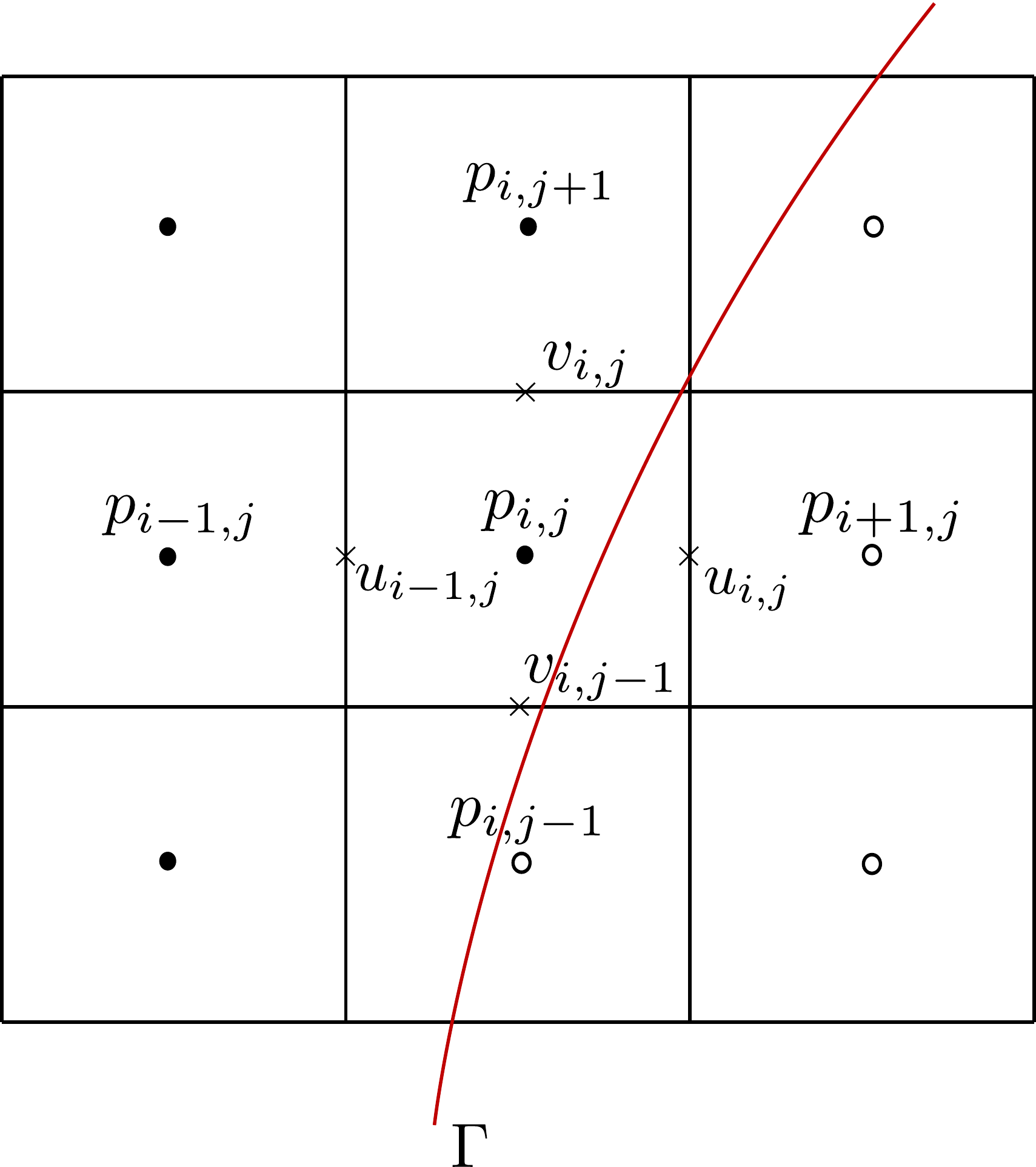}
 \end{center}
 \caption{Schematic of the 2D staggered grid where pressure locates at cell centers and velocity components locate at cell faces. The curved line specifies the interface $\Gamma$; filled and empty circles indicate discontinuous pressure (or density) values in phase 1 and 2, respectively.}
 \label{fig: GFM}
\end{figure}

As sketched in Fig.\ \ref{fig: GFM}, computing $\nabla^2 p$ at node $(i,j)$ requires three entries of $p$ in each direction. If CSF is used, all gradient terms can be evaluated with the straightforward central-difference, \ie
\begin{equation}
  \begin{aligned}
    (\nabla^2 p)_{i,j} = \frac{p_{i-1,j}^s -2p_{i,j}^s +p_{i+1,j}^s}{\Delta x^2}
                        +\frac{p_{i,j-1}^s -2p_{i,j}^s +p_{i,j+1}^s}{\Delta y^2}.
  \end{aligned}
  \label{laplace}
\end{equation}
However, the pressure at the cells adjacent to the interface will have to be smeared out; hence we denote them with $p^s$. In order for the pressure to be sharp, GFM creates an artificial fluid \Ge{(the ``ghost'' fluid)} and assumes that the discontinuity can be extended beyond the physical interface. That is, if we know the corresponding jumps of pressure, then its derivatives can be evaluated without smearing by removing such jumps. For the particular case depicted in Fig.\ \ref{fig: GFM}, Eq.\ \eqref{laplace} can be re-written as (see \cite{Liu_JCP_2000} for the intermediate steps)
\begin{equation}
  \begin{aligned}
    (\nabla^2 p)_{i,j} & =\frac{p_{i-1,j} -2p_{i,j} +p_{i+1,j}}{\Delta x^2} 
                         -\frac{[p]_{i,j}}{\Delta x^2} 
                         -\frac{1}{\Delta x} \bigg[\frac{\partial p}{\partial x} \bigg]_{i+1/2,j} \\
                       & +\frac{p_{i,j-1} -2p_{i,j} +p_{i,j+1}}{\Delta y^2}
                         -\frac{[p]_{i,j-1}}{\Delta y^2},
  \end{aligned}
  \label{laplace gfm}
\end{equation}
where we recall $[\cdot]_{i,j}$ denotes the discontinuity from fluid 1 to fluid 2 at cell $(i,j)$ (same for $[\cdot]_{i,j-1}$, \etc).

To determine the jump terms in Eq.\ \eqref{laplace gfm}, we first note that the velocity and its material derivatives across the interface of viscous flows are continuous \cite{Kang_JSC_2000, Desjardins_JCP_2008}, resulting in
\begin{equation}
    \bigg[\frac{1}{\rho^{n+1}} \nabla p^{n+1} \bigg]_\Gamma ={\bm 0}.
  \label{jump1}
\end{equation}
Furthermore, owing to the splitting that allows us to solve only for a constant-coefficient Poisson equation (Eq.\ \eqref{poisson split}), Eqs.\ \eqref{pressure split} and \eqref{jump1} lead to
\begin{equation}
    \bigg[ \frac{1}{\rho_0}\nabla p^{n+1} \bigg]_\Gamma + 
    \bigg[ (\frac{1}{\rho^{n+1}} -\frac{1}{\rho_0}) \nabla \hat{p} \bigg]_\Gamma = {\bm 0},
  \label{jump2}
\end{equation}
which also implies that the pressure \Ge{gradient} terms are continuous everywhere (\eg the subscript can be $(i+1/2,j)$), \Ge{along any direction}.

\Ge{Denoting the right-hand side of Eq.\ \eqref{poisson split} as $RP$, it is discretized as}
\begin{equation}
  \begin{aligned}
    RP_{i,j} &= 
    \bigg( \big(1-\frac{\rho_0}{\rho^{n+1}_{i+1/2,j}}) \frac{\partial \hat{p}}{\partial x}_{i+1/2,j}
    -      \big(1-\frac{\rho_0}{\rho^{n+1}_{i-1/2,j}}) \frac{\partial \hat{p}}{\partial x}_{i-1/2,j} \bigg) \big/\Delta x \\
    &+\bigg(\big(1-\frac{\rho_0}{\rho^{n+1}_{i,j+1/2}}) \frac{\partial \hat{p}}{\partial y}_{i,j+1/2}
    -       \big(1-\frac{\rho_0}{\rho^{n+1}_{i,j-1/2}}) \frac{\partial \hat{p}}{\partial y}_{i,j-1/2} \bigg) \big/\Delta y \\
    &- \frac{1}{\Delta x} \bigg[(1-\frac{\rho_0}{\rho^{n+1}}) \frac{\partial \hat{p}}{\partial x} \bigg]_{i+1/2,j} 
    +\frac{\rho_0}{\Delta t} \bigg( \frac{u^*_{i,j} - u^*_{i-1,j}}{\Delta x} + \frac{v^*_{i,j} - v^*_{i,j-1}}{\Delta y} \bigg),
  \end{aligned}
  \label{jump3}
\end{equation}
again using GFM \cite{Liu_JCP_2000}. \Ge{Comparing Eqs.\ \eqref{laplace gfm} and \eqref{jump3}, we note that the jump of the first derivatives cancels out recognizing Eq.\ \eqref{jump2}.} With a modified right-hand side, $RP^*$, defined as
\begin{equation}
  \begin{aligned}
    RP_{i,j}^* = RP_{i,j} + 
    \frac{1}{\Delta x} 
    \bigg[ (1-\frac{\rho_0}{\rho^{n+1}}) \frac{\partial \hat{p}}{\partial x} \bigg]_{i+1/2,j},
  \end{aligned}
  \label{jump4}
\end{equation}
the discrete form of Eq.\ \eqref{poisson split} reduces to
\begin{equation}
  \begin{aligned}
    \frac{p^{n+1}_{i-1,j}-2p^{n+1}_{i,j}+p^{n+1}_{i+1,j} }{\Delta x^2} + \frac{p^{n+1}_{i,j-1}-2p^{n+1}_{i,j}+p^{n+1}_{i,j+1} }{\Delta y^2} = 
    \frac{[p]^{n+1}_{i,j}}{\Delta x^2} + \frac{[p]^{n+1}_{i,j-1}}{\Delta y^2} + RP_{i,j}^*.
  \end{aligned}
  \label{gfm+split projection xy}
\end{equation}

Eq.\ \eqref{gfm+split projection xy} is still not ready to solve, since the pressure jumps for the first point away from the interface (\eg $[p]^{n+1}_{i,j}$) are not known. Following \cite{Desjardins_JCP_2008}, we perform a Taylor series expansion around $\Gamma$,
\begin{equation}
    [p]^{n+1}_{i,j} =  [p]^{n+1}_\Gamma + 
    (x_{i}-x_\Gamma)\bigg[\frac{\partial p}{\partial x} \bigg]^{n+1}_\Gamma +O((x_{i}- x_\Gamma)^2),
  \label{gfm taylor}
\end{equation}
where $[p]^{n+1}_\Gamma = \kappa_{\Gamma,x} /We$, and $\kappa_{\Gamma,x}$ is estimated from Eq.\ \eqref{2d icurv} in 2D and from Eq.\ \eqref{3d icurv2} in 3D, \Ge{along the $x$ direction} using $\phi^{n+1}_{i,j}$ and $\phi^{n+1}_{i+1,j}$. The jump of the pressure gradient at the interface can be similarly expanded at $(i,j)$
\begin{equation}
    \bigg[\frac{\partial p}{\partial x} \bigg]^{n+1}_\Gamma = 
    \bigg[\frac{\partial p}{\partial x} \bigg]^{n+1}_{i,j} +O(x_\Gamma -x_{i}),
  \label{gfm taylor 2}
\end{equation}
resulting in
\begin{equation}
    [p]^{n+1}_{i,j} =  \frac{\kappa_{\Gamma,x}}{We} + 
    (x_{i}-x_\Gamma)\bigg[\frac{\partial p}{\partial x} \bigg]^{n+1}_{i,j} +O((x_{i}- x_\Gamma)^2).
  \label{gfm taylor 3}
\end{equation}

Using Eq.\ \eqref{jump2}, we can re-write Eq.\ \eqref{gfm taylor 3} as
\begin{equation}
    [p]^{n+1}_{i,j} =  \frac{\kappa_{\Gamma,x}}{We} + (x_{i}-x_\Gamma) 
    \bigg[ (1-\frac{\rho_0}{\rho^{n+1}}) \frac{\partial \hat{p}}{\partial x} \bigg]_{i,j} +O((x_{i}- x_\Gamma)^2),
  \label{gfm taylor 4}
\end{equation}
where the jump term on the right-hand side can be explicitly calculated using the family of identities \Ge{of the form \cite{Kang_JSC_2000}}
\begin{equation}
    \Ge{[AB]=[A]\tilde{B}+\tilde{A}[B],\quad \tilde{A}=aA_1+bA_2,\quad a+b=1.}
  \label{jump identity}
\end{equation}

%
%

Although Eqs.\ \eqref{gfm taylor 4} and \eqref{jump identity} lead to a second-order pressure jump, it is much simpler to keep only the leading-order term, \ie
\begin{equation}
    [p]^{n+1}_{i,j} =  \frac{\kappa_{\Gamma,x}}{We} + O(x_{i}- x_\Gamma).
  \label{gfm taylor 6}
\end{equation}
This way, the pressure jump varies only with the local curvature, remains invariant across the interface, and is second-order accurate when the density is uniform. For the test cases shown below, Eq.\ \eqref{gfm taylor 6} is used. Thus, the complete discretization of Eq.\ \eqref{poisson split} reads
\begin{equation}
  \begin{aligned}
    \frac{p^{n+1}_{i-1,j}-2p^{n+1}_{i,j}+p^{n+1}_{i+1,j} }{\Delta x^2} + \frac{p^{n+1}_{i,j-1}-2p^{n+1}_{i,j}+p^{n+1}_{i,j+1} }{\Delta y^2} = 
    \frac{1}{We} \bigg( \frac{\kappa_{\Gamma,x}}{\Delta x^2}+\frac{\kappa_{\Gamma,y}}{\Delta y^2} \bigg)
    + RP_{i,j}^*,
  \end{aligned}
  \label{gfm+split projection xy 1}
\end{equation}
\Ge{with $RP^*_{i,j}$ defined in Eq.\ \eqref{jump4} corresponding to Fig.\ \ref{fig: GFM}.}

Clearly, the resulting linear system (Eq.\ \eqref{gfm+split projection xy 1}) has a standard \Ge{positive definite,} symmetric coefficient matrix, and it can be solved directly using the FFT-based fast Poisson solver (Sec.\ \ref{p correction}). Care should be exercised when a nodal point crosses the interface in more than one direction. In those cases, \Ge{the interface curvature of each crossing direction may be different and it shall not be averaged. Otherwise, the projection (Eq.\ \eqref{poisson split}) and correction (Eq.\ \eqref{projection 1}) steps can become inconsistent, making the velocity not divergence-free.} Additionally, when taking the gradient of the pressure-correction term; \eg its derivative along the $x$ direction, the correct discretization should be
\begin{equation}
  \begin{aligned}
    \frac{\partial \hat{p}}{\partial x}_{i,j} = 
    \frac{\big(\hat{p}_{i+1,j}-(2[p]^{n}_{i+1,j}-[p]^{n-1}_{i+1,j}) \big) - \hat{p}_{i,j}}{\Delta x}. 
  \end{aligned}
  \label{gfm grad}
\end{equation}

\noindent After removing the jump, the divergence of the bracket term in Eq.\ \eqref{poisson split} is evaluated in the same way as in \cite{Dodd_JCP_2014}.

Finally, we can re-write Eqs.\ \eqref{poisson split} and \eqref{projection 1} compactly as
\begin{equation}
  \begin{aligned}
    \nabla ^2 p^{n+1} = \nabla ^2_g [p]_\Gamma + \nabla \cdot \bigg[ \big(1-\frac{\rho_0}{\rho^{n+1}}) \nabla_g \hat{p} \bigg] + \frac{\rho_0}{\Delta t} \nabla \cdot {\bm u}^*,
  \end{aligned}
  \label{gfm+split projection}
\end{equation}
\begin{equation}
    {\bm u}^{n+1} = {\bm u}^* -\Delta t \bigg[\frac{1}{\rho_0} \nabla_g p^{n+1} + \big(\frac{1}{\rho^{n+1}} - \frac{1}{\rho_0}\big) \nabla_g \hat{p} \bigg].
  \label{gfm+split correction}
\end{equation}
\noindent where $\nabla_g$ and $\nabla ^2_g [p]_\Gamma$ denote, respectively, the gradient operator considering the jump and the extra jump terms from the laplacian operator due to GFM.


\subsection{Time integration}
\label{sec: temporal}

In the current work, a second-order accurate Adams-Bashforth scheme is used for the time integration. The time step is restricted by convection, diffusion, surface tension, and gravity, due to our explicit treatment of these terms. As suggested in \cite{Kang_JSC_2000}, the overall time step restriction is

\begin{equation}
  \Delta t \leqslant 1\bigg/ \bigg(C_{CFL}+V_{CFL}+\sqrt{(C_{CFL}+V_{CFL})^2+4G_{CFL}^2+4S_{CFL}^2}\bigg),
  \label{dt}
\end{equation}

\noindent where $C_{CFL}$, $V_{CFL}$, $G_{CFL}$, and $S_{CFL}$ are the ``speeds" due to convection, viscosity, gravity, and surface tension, respectively. Specifically, they are given as

\begin{equation}
  C_{CFL} = \frac{|u|_{max}}{\Delta x}+\frac{|v|_{max}}{\Delta y}+\frac{|w|_{max}}{\Delta z},
  \label{dt conv}
\end{equation}

\begin{equation}
  V_{CFL} = \frac{1}{Re} max \bigg(\frac{\mu_1}{\rho_1},\frac{\mu_2}{\rho_2} \bigg) \bigg(\frac{2}{\Delta x^2}+\frac{2}{\Delta y^2}+\frac{2}{\Delta z^2}\bigg),
  \label{dt visc}
\end{equation}

\begin{equation}
  G_{CFL} = \sqrt{\frac{1}{Fr}\frac{|(1-\tfrac{\rho_1+\rho_2}{2\rho}) g|_{max}}{min (\Delta x,\Delta y,\Delta z)} },
  \label{dt grav}
\end{equation}

\begin{equation}
  S_{CFL} = \sqrt{\frac{1}{We}\frac{|\kappa|_{max}}{min(\rho_1,\rho_2) \big[min(\Delta x,\Delta y\Delta z)\big]^2}}.
  \label{dt surf}
\end{equation}

\noindent where $|\kappa|_{max}$ in \eqref{dt surf} can be approximated by $1/\Delta x$ in 2D and $2/\Delta x$ in 3D, assuming $\Delta x$ is the smallest grid spacing.

The reasons we choose an explicit temporal scheme rather than an implicit one are twofold. First, for applications involving a large density and viscosity contrast, the stability restriction imposed by surface tension is usually greater than that imposed by diffusion. Second, an implicit formulation of GFM has been admitted to be challenging to develop \cite{Desjardins_JCP_2008}, and it was shown in a recent study \cite{Denner&Wachem_JCP_2015} that a capillary time-step constraint exists, irrespective of the type of implementation, due to the temporal sampling of surface capillary waves. Fortunately, the fast pressure-correction method enables the use of FFT for the constant-coefficient Poisson equation and hence an accurate and fast solution of the two-fluid Navier-Stokes equation can be obtained.


\subsection{Full solution procedure}
\label{sec: procedure}

We summarize the full solution procedure as follows:

\begin{enumerate}

    \item Advance the interface explicitly from $\phi^{n}$ to $\phi^{n+1}$ using the ICLS, and update the density $\rho^{n+1}$ and the viscosity $\mu^{n+1}$.
    
    \item Advance the velocity field explicitly from ${\bm u}^{n}$ to ${\bm u}^{*}$ with Eqs.\ \eqref{ru} and \eqref{ab2}.
    
    \item Project the velocity field by solving the constant-coefficient Poisson Eq.\ \eqref{gfm+split projection} making use of the FastP* and the GFM.
    
    \item Update the velocity from ${\bm u}^{*}$ to ${\bm u}^{n+1}$ explicitly with Eq.\ \eqref{gfm+split correction}, again using the FastP* and the GFM.

\end{enumerate}


\subsection{Validations}
\label{sec: validations}

\Ge{In this section, we validate the coupled ICLS/NS solver using three benchmark examples with increasing complexities. Specifically, the first example verifies the discrete momentum balance for fluids of the same density and viscosity. This concerns the surface tension computed by the GFM using interface curvatures. Then, the density and viscosity ratios are significantly increased (up to $10^4$) to test the combined FastP* and GFM. Using the same test, we also provide a convergence check of the complete flow solver. Finally, the overall accuracy is assessed by simulating a 3D bubble in comparison with experiments.}

\subsubsection{Spurious currents}
\label{sec: spurious}

A common problem in multiphase-flow simulations is the artificial velocity generated at the fluid interface due to errors in the curvature computation. To access the significance of such spurious currents, we test a stationary droplet of diameter $D=0.4$ placed at the center of a unit box. The surface tension between the inner and outer fluid is $\sigma=1$, the viscosity is uniformly $\mu=0.1$, and the density ratio is 1. By changing the density $\rho$ of both fluids, the Laplace number $La=\sigma \rho D /\mu^2$ can be varied. The spurious currents are thus determined from the resulting capillary number $Ca=\abs{U_{max}} \mu / \sigma$ at a non-dimensional time $t\sigma/(\mu D)=250$. Here, we compare the results on a $32\times32$ mesh with the GFM implementation by Desjardins \etal \cite{Desjardins_JCP_2008}. As listed in Table \ref{tab: spurious}, the capillary numbers from both tests remain very small for all the Laplace numbers, with the present results being one-order smaller. 

\begin{table}[t]
 \centering
  \caption{Dependence of spurious current capillary number $Ca$ on the Laplace number for a static droplet with surface tension on a $32\times32$ mesh in comparison with Desjardins \etal \cite{Desjardins_JCP_2008}.}
  \tabulinesep=1.2mm
  \begin{tabular}{ l l l l l l l}
   \hline
   La  &12 &120 &1,200 &12,000 &120,000 &1,200,000    \\
   \hline
   Ca  &2.85\e{-6}&3.14\e{-6}&3.63\e{-6}&3.87\e{-6}&3.41\e{-6}&5.79\e{-7}\\
   Ca from \cite{Desjardins_JCP_2008}&4.54\e{-5}&3.67\e{-5}&3.62\e{-5}&4.15\e{-5}&3.75\e{-5}&8.19\e{-6}\\
   \hline
   \label{tab: spurious}
  \end{tabular}
\end{table}

\begin{figure}[t]
\centering
  \includegraphics[width=.5\columnwidth]{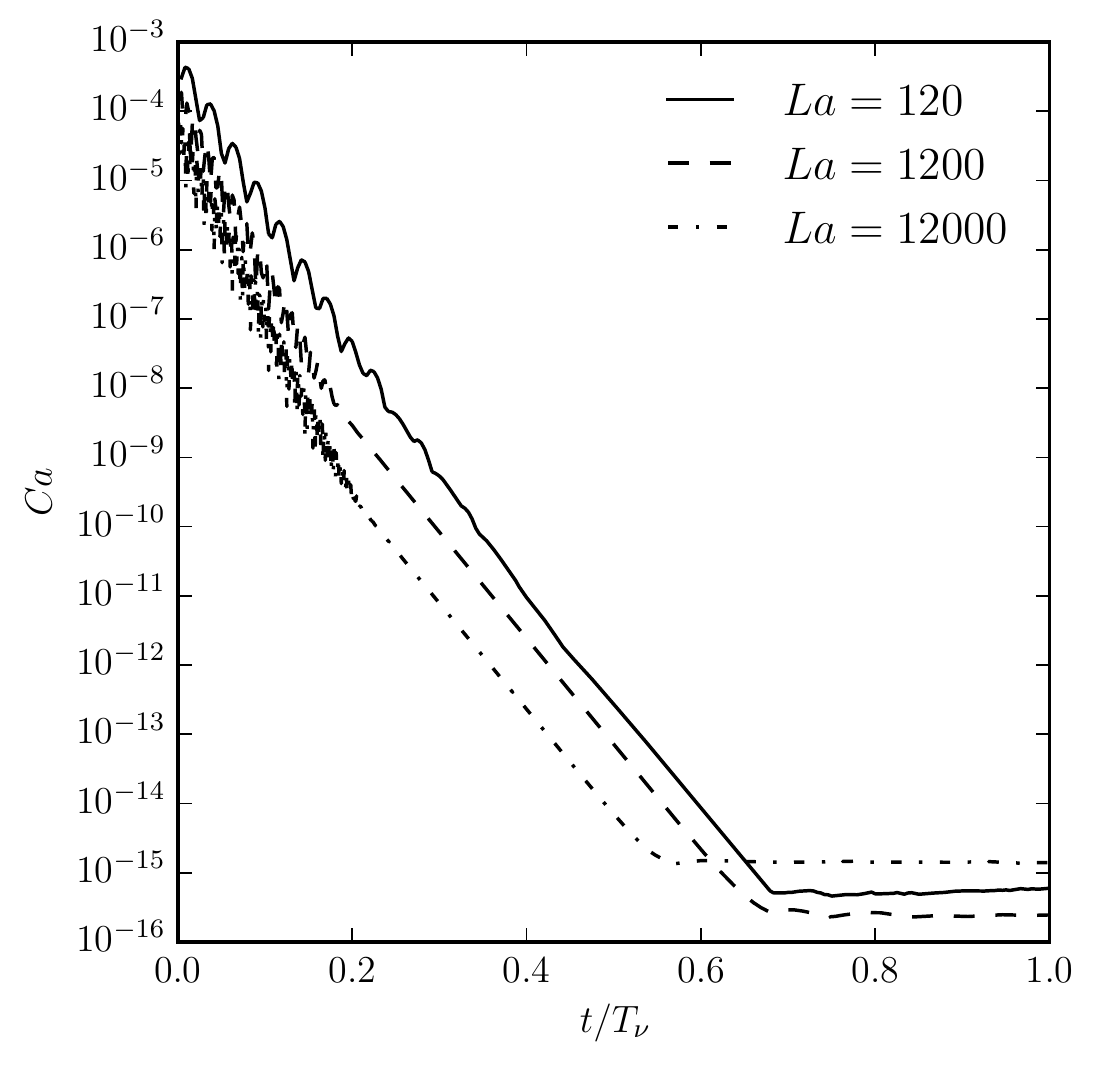}
   \caption{Temporal evolution of the spurious currents without performing level set reinitialization at three Laplace numbers as in \cite{Popinet_JCP_2009}.}
   \label{fig: zero spurious}
\end{figure}

We also note that the spurious currents reported in Table \ref{tab: spurious} are obtained by performing the level set reinitialization at about every 100 time steps. However, if we turn off the reinitialization, such spurious velocity will eventually go to machine zero, as shown in Fig.\ \ref{fig: zero spurious}, where time is non-dimensionalized with the viscous time scale, $T_{\nu}=\rho D^2/\mu$. The nearly exponential decay of $Ca$ and the collapsing of the three curves are the result of the viscous damping of the spurious velocity, as the shape of droplet relaxes to its numerical equilibrium. Similar results are obtained and explained in greater detail in \cite{Popinet_JCP_2009} using a balanced-force continuum-surface-force surface-tension
formulation and the VOF. 
The result in Fig.\ \ref{fig: zero spurious} therefore validates the computation of the surface tension with the GFM.

\subsubsection{Capillary wave}
\label{subsec: cap_wave}

To verify the solver at large density and (dynamic) viscosity contrasts, we simulate a small-amplitude capillary wave for which there exists an analytical solution derived by Prosperetti \cite{Prosperetti_1981}. Specifically, \Ge{an initially sinusoidal interface is imposed between two immiscible, viscous fluids of infinite depth and lateral extent. When the lower fluid is heavier}, the balance between inertia, viscosity, and surface tension results in a decaying free-surface wave. By requiring matching kinematic viscosity $\nu_u=\nu_l$ \Ge{($u$ for upper, $l$ for lower)}, the solution of the wave amplitude \Ge{in terms of Laplace transforms can be inverted} analytically and compared with the simulation results. 

We set up our simulation in the same way as suggested in \cite{Dodd_JCP_2014}. Here, two fluids of equal depth are placed in a $1\times3$ ($64\times192$ grid points) domain, where the streamwise direction ($L=1$) is periodic and the vertical direction ($H=3$) wall-bounded. The interface has an initial wavelength of $\lambda=1$ and an amplitude of $a_0=0.01$. With varying density ratios $\rho_l/\rho_u$, the non-dimensional parameters for the test are
\begin{equation}
    Re=100, \quad We=1, \quad Fr=\infty, \quad \rho_l/\rho_u=10-10,000, \quad \nu_l=\nu_u.
  \label{capillary_wave_test}
\end{equation}
The CFL number $\Delta t/\Delta x$ is $2.5\e{-2}$ for $\rho_l/\rho_u=10$ and $10^2$, and it is reduced to $2.5\e{-3}$ for $\rho_l/\rho_u=10^3$ and $2.5\e{-4}$ for $\rho_l/\rho_u=10^4$ .

\begin{figure}[t]
\centering
  \subfigure[$\rho_l/\rho_u=10$, $\nu_l=\nu_u$]{\includegraphics[width=.45\columnwidth]{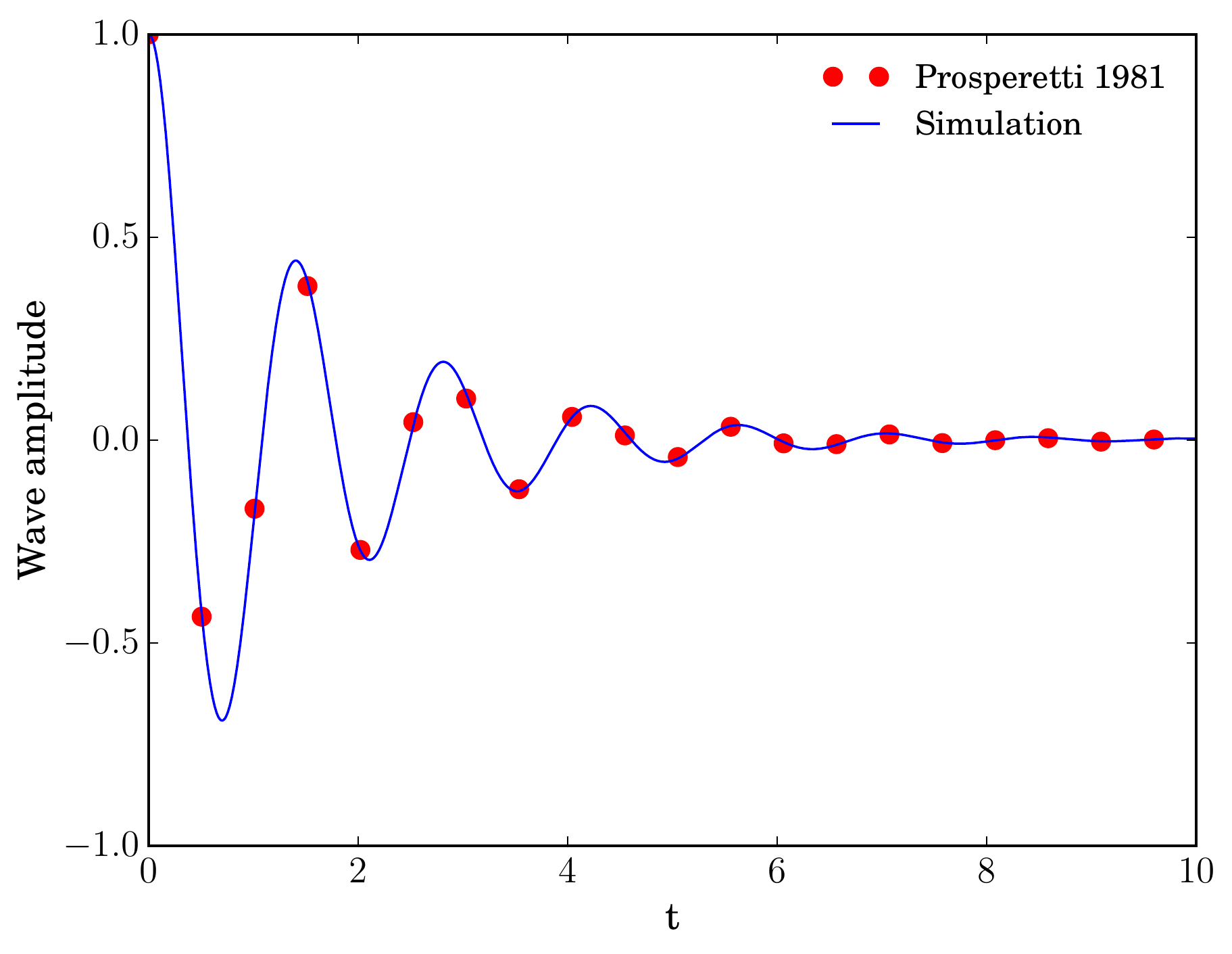}}
  \subfigure[$\rho_l/\rho_u=100$, $\nu_l=\nu_u$]{\includegraphics[width=.45\columnwidth]{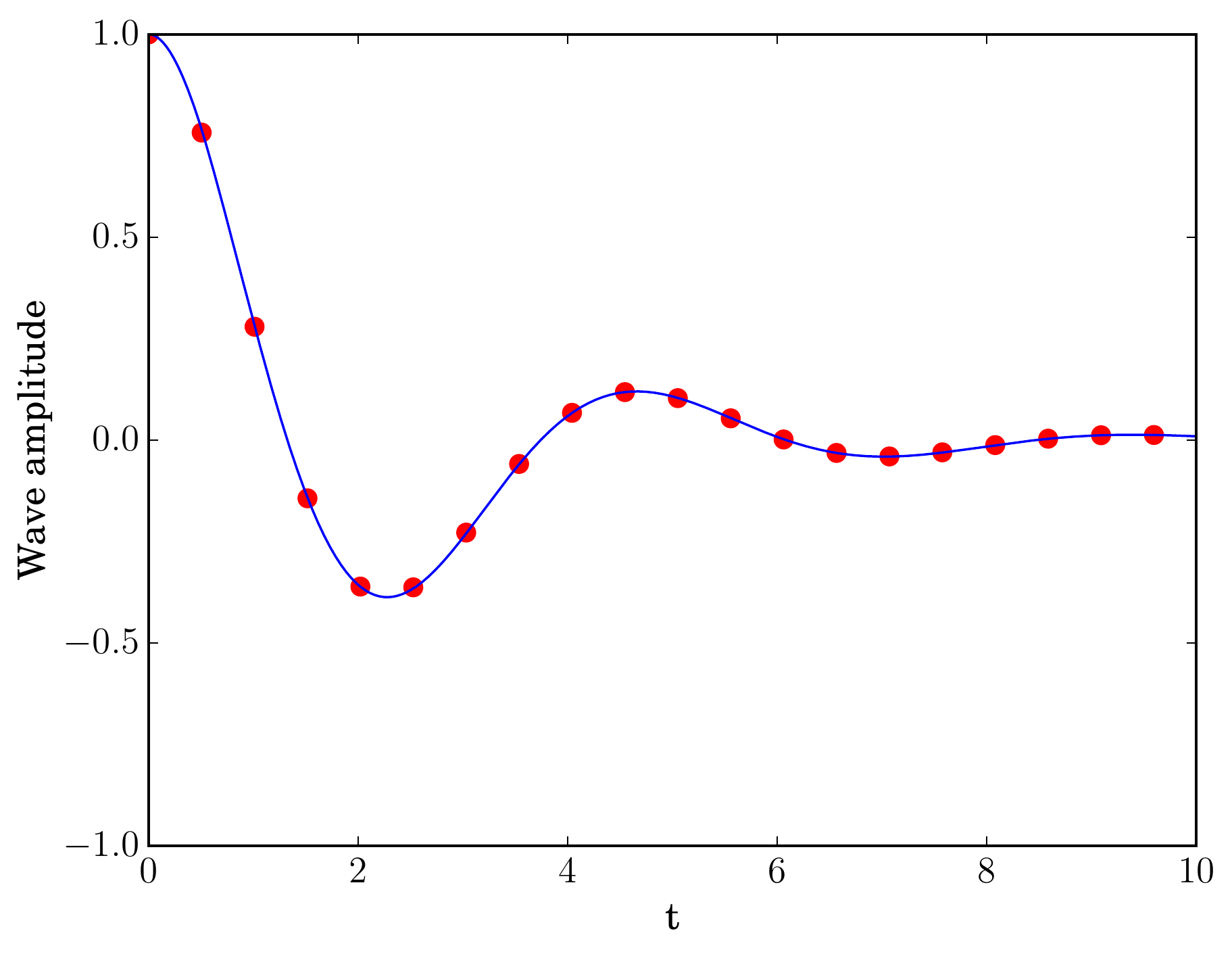}}
  \subfigure[$\rho_l/\rho_u=1000$, $\nu_l=\nu_u$]{\includegraphics[width=.45\columnwidth]{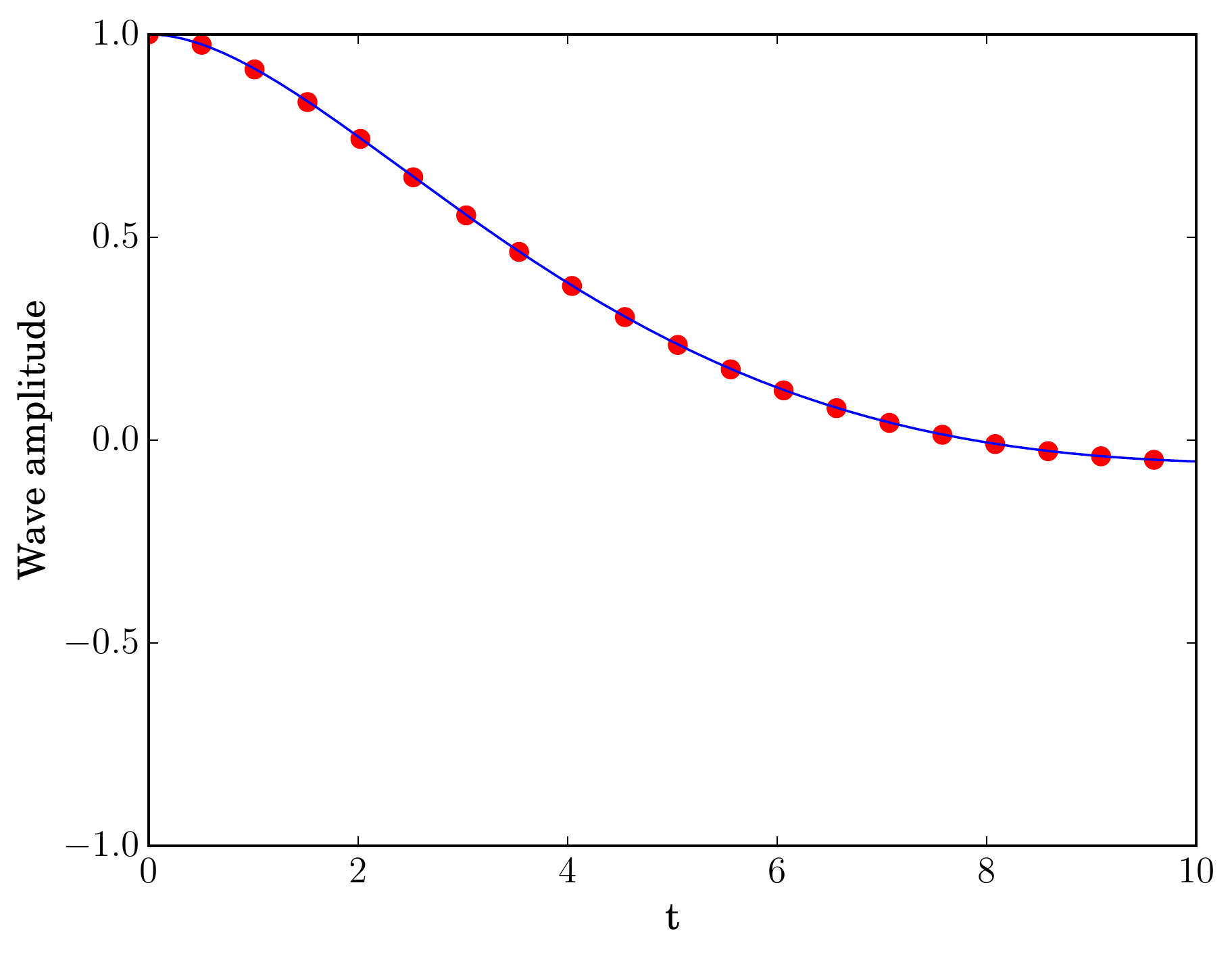}}
  \subfigure[$\rho_l/\rho_u=10000$, $\nu_l=\nu_u$]{\includegraphics[width=.45\columnwidth]{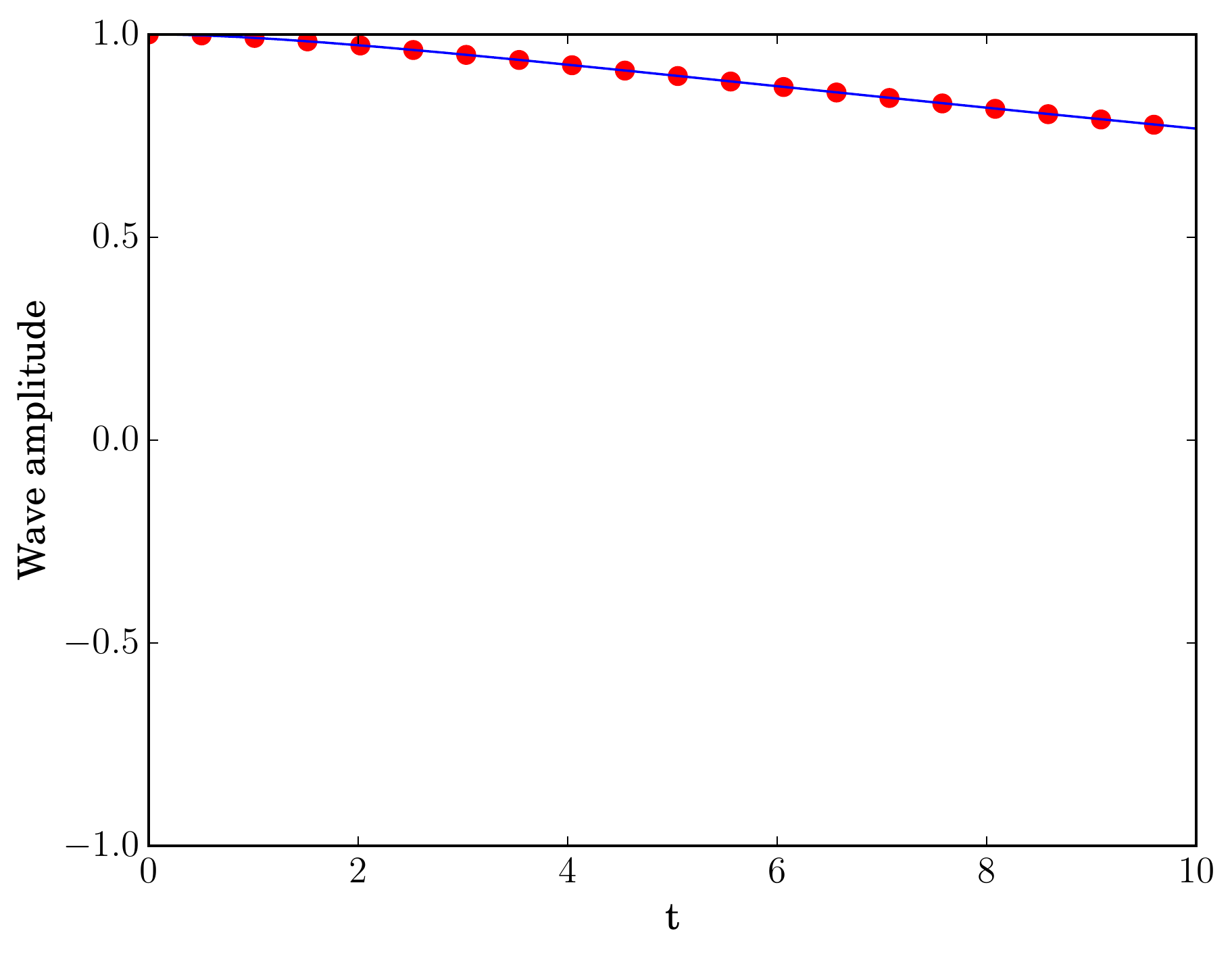}}
   \caption{Time development of the capillary wave amplitude (normalized to $a_0$) for increasing density ratios and matching kinematic viscosity in comparison with Prosperetti's analytical solution \cite{Prosperetti_1981}.}
   \label{fig:cap_wave}
\end{figure}

Fig.\ \ref{fig:cap_wave} shows the temporal evolution of the wave amplitude up to $t=10$. The excellent agreement with Prosperetti's analytical solution \cite{Prosperetti_1981} confirms the normal stress balance computed using the GFM. \Ge{And accurate results at very large density contrasts are realized by combining the FastP* with GFM.} Note that the dynamic viscosity ratio $\mu_l/\mu_u$ also varies from $10$ to $10^4$. However, neglecting its contribution to the pressure jump by regularizing the viscosity profile yields accurate results since the Capillary number is small ($Ca=We/Re=0.01$), as discussed in conjunction with Eq.\ \eqref{pressure jump 2}.


\subsubsection{\Ge{Convergence}}

\Ge{We continue to check the temporal and spatial convergence rates of the coupled ICLS/NS solver. Here, the same test problem as in Sec.\ \ref{subsec: cap_wave} is used, with the non-dimensional parameters given as}
\begin{equation}
    \Ge{Re=500, \quad We=1, \quad Fr=\infty, \quad H_0=0.05, \quad \rho_l/\rho_u=20, \quad \mu_l/\mu_u=20,}
  \label{convergence test}
\end{equation}
\Ge{again following \cite{Dodd_JCP_2014}. Placing the fluids in a $1 \times 1$ box, the flow is simulated under different time steps or on different meshes so that the errors can be computed between successive solutions.}

\begin{table}[t]
    \centering
    \caption{\Ge{Temporal and spatial convergence rates for the velocity component $u$ and the pressure $p$.}}
    \tabulinesep=1.2mm
    \begin{tabular}{ l l l l l l l}
      \hline
      &$L_2^{4\Delta t,2\Delta t}$&$L_2^{2\Delta t,\Delta t}$&Rate\quad\quad&$L_2^{4\Delta x,2\Delta x}$&$L_2^{2\Delta x,\Delta x}$&Rate\\
      \hline                                                                                                             
      $u$  &2.46\e{-8}            &1.03\e{-8}         &1.19 &2.16\e{-7}           &5.95\e{-8}        &1.82 \\
      $p$  &1.13\e{-6}            &3.85\e{-7}         &1.46 &3.25\e{-3}           &6.11\e{-4}        &2.67 \\
     \hline
 \end{tabular}
 \label{tab: convergence}
\end{table}

\Ge{Table \ref{tab: convergence} shows the convergence rates for the velocity component $u$ and the pressure $p$ in the $L_2$ norm. Here, the temporal convergence is evaluated at $t=6.25\e{-2}$ on a $256^2$ grid, by increasing the time step from $\Delta t=4.88\e{-5}$ to $2\Delta t$ and $4\Delta t$. Two iterations of reinitialization are performed every $25-100$ time steps. The observed convergence rates for both velocity and pressure is between first and second order. Considering that we use RK3 for LS and AB2 for NS, the reduced convergence is probably due to the reinitialization that perturbs the interface. Changing the frequency of the reinitialization, we indeed observe different convergence rates (they can also exceed second order if the density ratio is 1, not shown). Next, the spatial convergence is obtained by successively refining the grid from $32^2$ to $64^2$ to $128^2$. Using the same time step $\Delta t=4.88\e{-5}$ and interpolating the solution to the coarse grid after one solve, the results display nearly second order convergence for the velocity and a super-convergence for the pressure. We note that the GFM has been proven convergent (but without a rate) for variable-coefficient Poisson equations \cite{Liu_MC_2003}. Our results thus show improved accuracy in two fluid problems, when a constant-coefficient Poisson equation is obtained by combining the GFM with the FastP*.}


\subsubsection{Rising bubble}

Finally, we compute four cases of a rising bubble to access the overall accuracy of the current ICLS/NS solver in 3D in the presence of moderate deformations. Originally documented by Grace \cite{Grace_1973}, it was observed that a single gas bubble rising in quiescent liquid has four characteristic shapes: spherical, ellipsoidal, skirted, or dimpled. The governing non-dimensional numbers are the Morton number $M$, Eotvos number $Eo$ \Ge{(sometimes referred to as the Bond number)}, and the terminal Reynolds number $Re_t$, defined as
\begin{equation}
  \begin{aligned}
    M=\frac{g \mu_l^4}{\rho_l\sigma^3}, \quad Eo=\frac{\Delta \rho g d^2}{\sigma}, \quad Re_t=\frac{\rho_l U_{\infty}d}{\mu_l},
  \end{aligned}
  \label{bubble non-di}
\end{equation}
\noindent where $d$ is the bubble diameter, $\Delta \rho$ is the density difference, $U_\infty$ is the terminal velocity of the bubble, and the subscripts $l$ and $g$ denote, in order, the liquid and gas phase. The Morton and Eotvos number are defined purely by the material properties of the chosen fluids, while the terminal Reynolds number provides a measure of the steady-state bubble velocity.

Table \ref{tab: bubble} lists the four representative cases we select for the simulations. A spherical bubble of diameter $d=1$ is centered in a domain of size $(L_x\times L_y \times L_z) = (3d\times6d\times3d)$. 
A grid of $96\times192\times96$ points is used, giving the bubble an initial resolution of 32 points per diameter. 
Periodic boundary conditions are imposed in the $x$ (spanwise) and $y$ (rising) directions whereas no friction, no penetration is enforced in the $z$ direction. As suggested by Annaland \etal \cite{Annaland_2005}, a ratio of 100 
between the density and viscosity of liquid and gas is sufficiently high to approximate such gas-liquid systems, leading to $\Delta \rho \approx \rho_l$. $Re$ and $We$ in Eq.\ \eqref{non-di} can thus be obtained from $M$ and $Eo$ as
\begin{equation}
  \begin{aligned}
    Re=\bigg(\frac{Eo^3}{M} \bigg)^{1/4}, \quad We=Eo.
  \end{aligned}
\end{equation}

\noindent The CFL number, $\Delta t/\Delta x$, is $1.6\e{-4}$ for cases (a), (b), and (d), and $1.6\e{-3}$ for case (c). The simulation is integrated in time up to $t=10$ to ensure the bubble reaches nearly steady state.

\begin{figure}[t]
    \centering
    \subfigure[Spherical]{\includegraphics[width=.24\columnwidth]{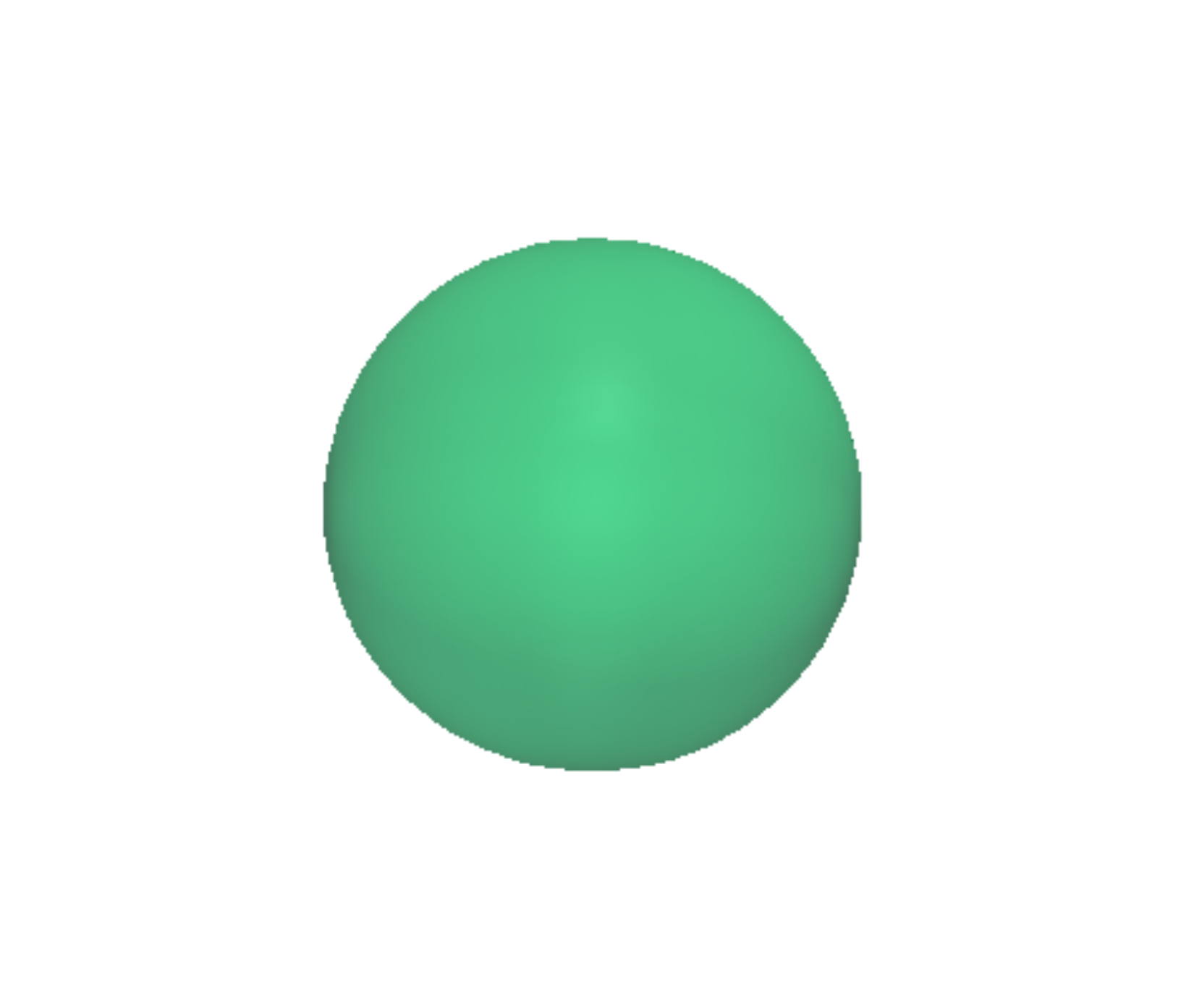}}
    \subfigure[Ellipsoidal]{\includegraphics[width=.24\columnwidth]{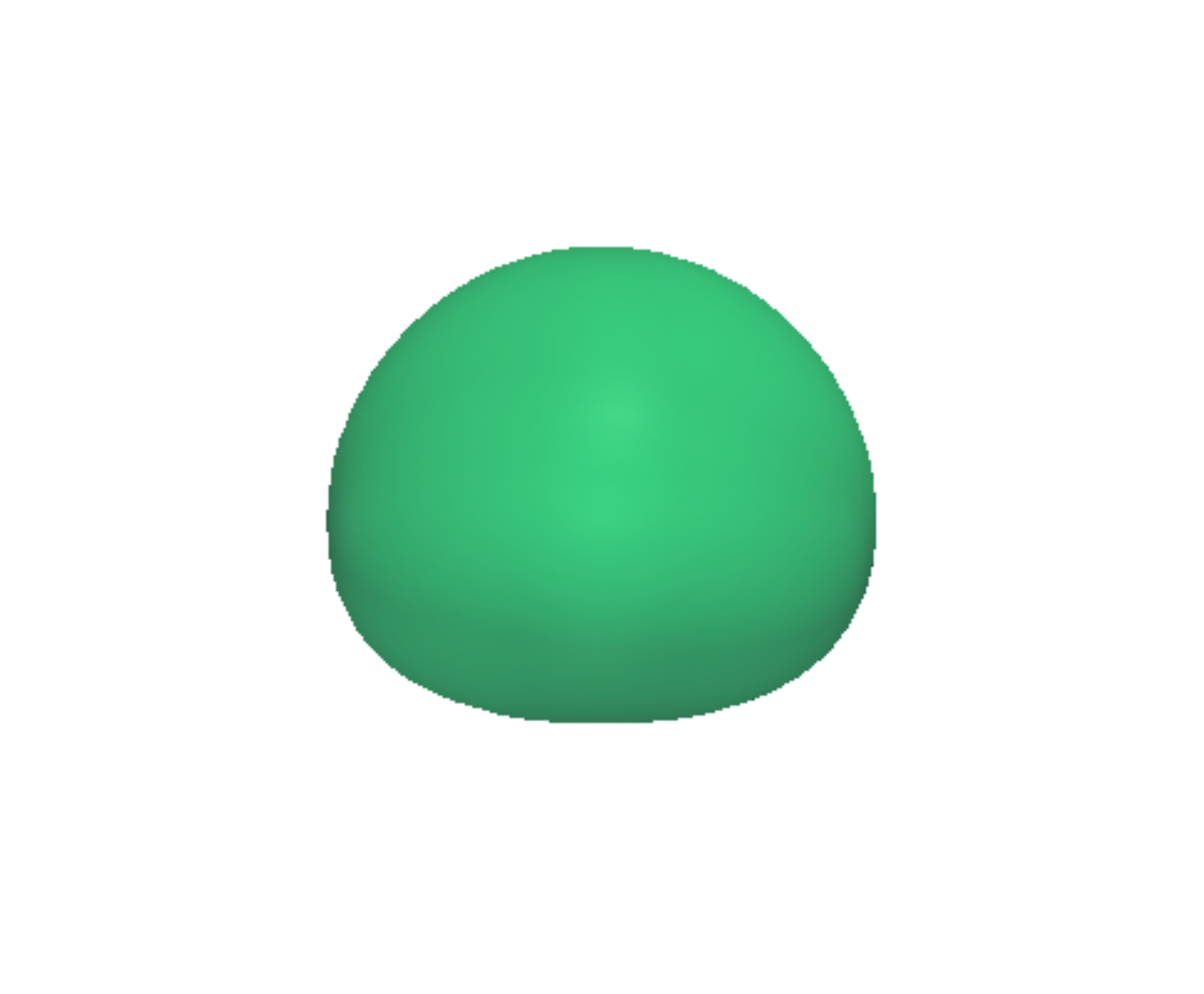}}
    \subfigure[Skirted]{\includegraphics[width=.24\columnwidth]{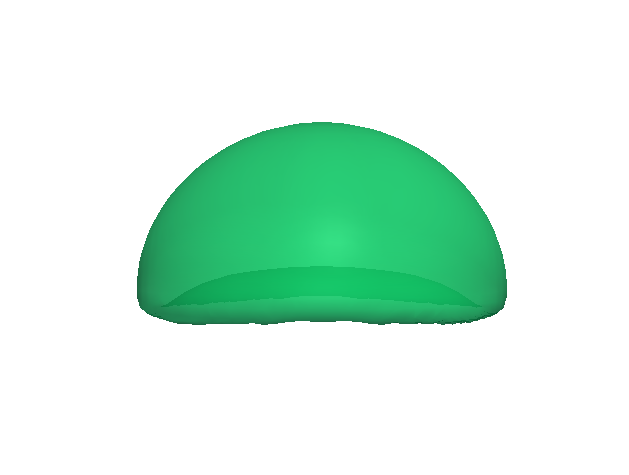}}
    \subfigure[Dimpled]{\includegraphics[width=.24\columnwidth]{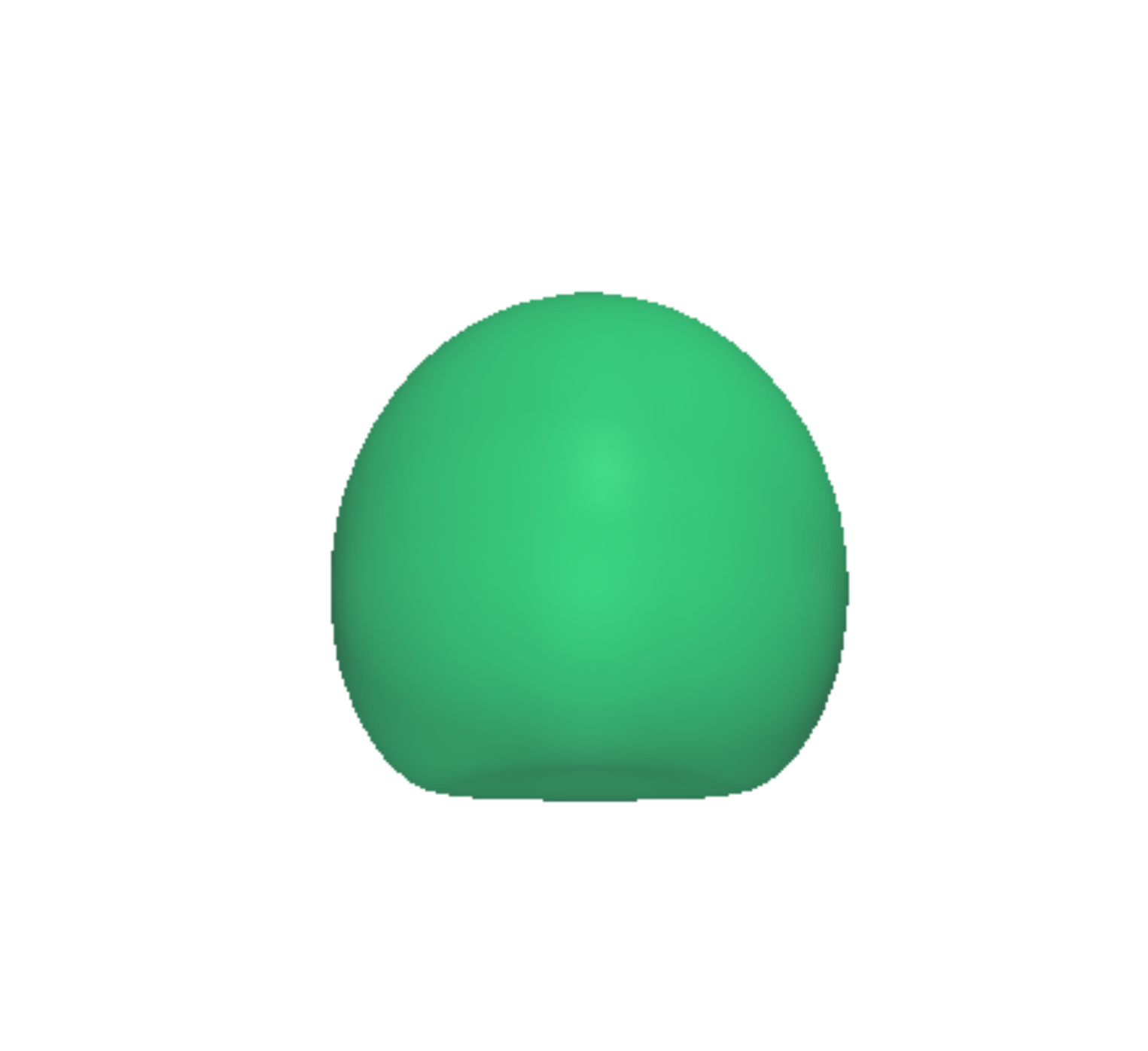}}
    \caption{Bubble shapes resulting from different Morton (M) and Eotvos (Eo) numbers, as indicated in Table \ref{tab: bubble}.}
    \label{fig: bubble shape}
\end{figure}

The results of the bubble terminal velocities are presented in Table \ref{tab: bubble}. The difference between the computed Reynolds, $Re_C$, and the terminal Reynolds, $Re_G$, measured by Grace \cite{Grace_1973} remains small for all four cases. The bubble mass is conserved, with a maximal mass loss of about $0.02\%$ found in the skirted case, where the bubble undergoes a large and rapid deformation. The corresponding bubble shapes are illustrated in Fig.\ \ref{fig: bubble shape}, which clearly displays spherical, ellipsoidal, skirted, and dimpled shapes. We can therefore conclude that the dynamics of a single rising bubble is well-captured.

\begin{table}[t]
 \centering
  \caption{Comparison of computed terminal Reynolds number ($Re_C$) and experimental terminal Reynolds number ($Re_G$) obtained from the Grace diagram \cite{Grace_1973} under four different Morton (M) and Eotvos (Eo) numbers.}
  \tabulinesep=1.2mm
   \begin{tabular}{ l l l l l l c }
      \hline
      Case  & Bubble regime   &$M$        &$Eo$      &$Re_G$     &$Re_C$     &Mass loss (\%) \\
      \hline
      (a)   &Spherical        &1\e{-3}    &1         &1.7        &1.73        &9.86\e{-5}  \\
      (b)   &Ellipsoidal      &0.1        &10        &4.6        &4.57        &3.32\e{-4}  \\
      (c)   &Skirted          &1          &100       &20.0       &19.21       &1.64\e{-2}  \\
      (d)   &Dimpled          &1000       &100       &1.5        &1.71        &3.28\e{-3}  \\
      \hline
   \label{tab: bubble}
  \end{tabular}
\end{table}

\section{Droplet interactions}
\label{sec: idrop}

\Ge{A unique feature of colloidal suspensions is the interaction between neighboring droplets, displaying fascinating behaviors such as self-assembly, self-replication, \etc. The reason for such interactions is rather complex; it often arises from a combination of fluid mechanical effects and physicochemical properties of the substance. To study the droplet interactions in the present ICLS/NS framework, we provide in this section a hydrodynamic model for the depletion forces. The method is a natural extension of the LS and GFM, and we demonstrate the clustering of droplets in various structures from a dumbbell to a face-centered cubic crystal.}

\subsection{Extension to multiple level set}
\label{subsec: mls}

\Ge{The level set method discussed so far involves one marker function; we call it single level set (SLS) method. Thanks to its Eulerian nature, SLS can describe many droplets at the same time, provided that they do not need to be distinguished from each other. On the other hand,} 
SLS can also be extended to multiple level set (MLS), so that each droplet has its own color function. This has several benefits including distinction and tracking of each droplet, independent curvature computation, and ability to prevent numerical coalescence, \etc. Furthermore, with the narrow band approach \cite{Adalsteinsson_JCP_1995,Peng_JCP_1999} \Ge{and the various other techniques introduced in Sec.\ \ref{intro} \cite{Nielsen_JSC_2006,Brun_JCP_2012}}, the additional computational \Ge{and memory} cost as the number of the level set functions increases is limited.


\begin{figure}[t]
 \begin{center}
 \includegraphics[width=5cm]{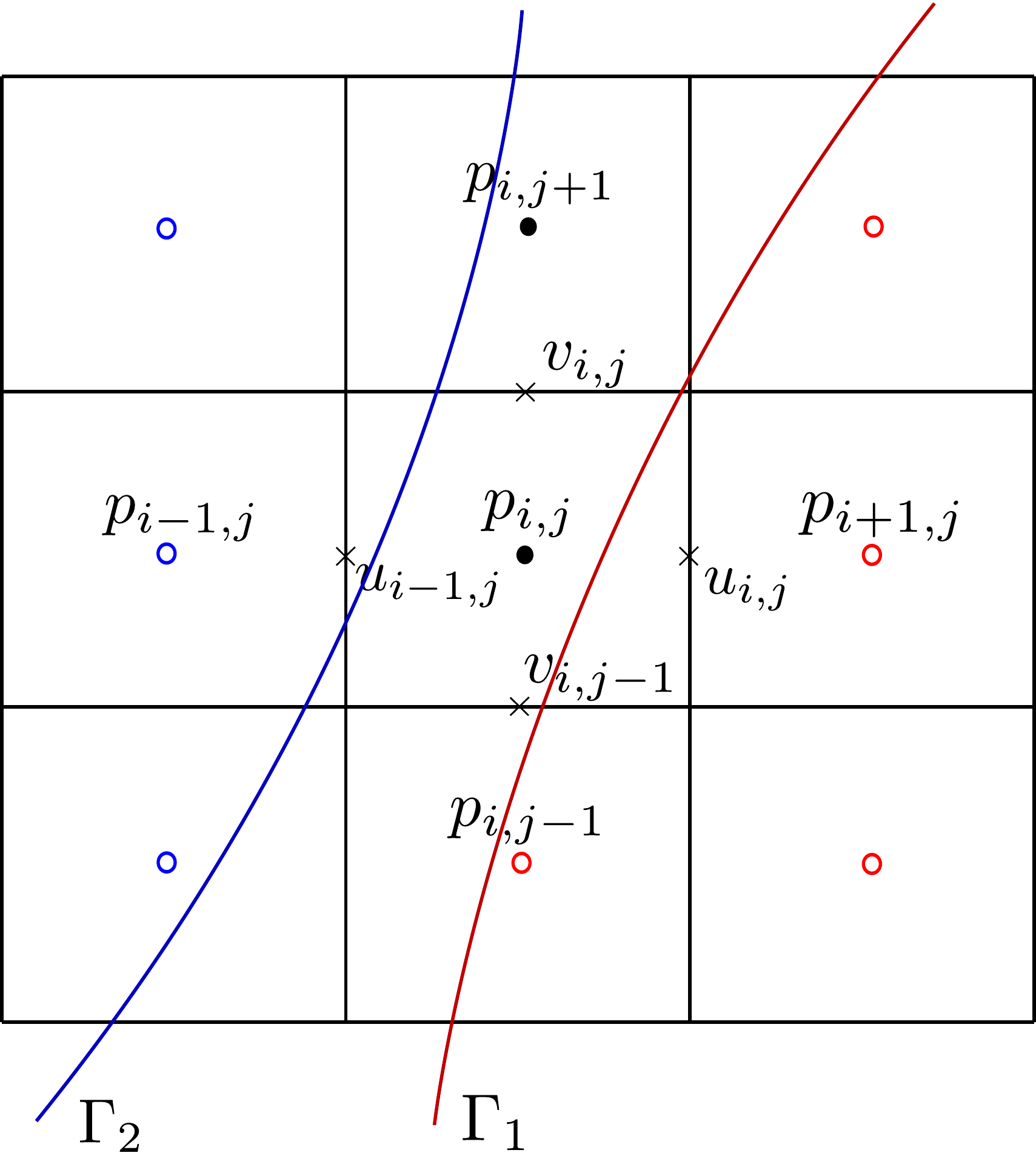}
 \end{center}
 \caption{Pressure jump in the presence of multiple interfaces within two grid cells. Red and blue circles indicate nodal pressure in droplet 1 and 2, respectively. (For interpretation of the references to color in this figure legend, the reader is referred to the web version of this article.)}
 \label{fig: GFM close}
\end{figure}

The extension from SLS to MLS is straightforward. Assuming no droplets will overlap, each level set function is simply advected successively. When two droplets get close (typically within two grid cells, see Fig.\ \ref{fig: GFM close}), the pressure jump across each interface needs to be considered and superimposed. That is, Eq.\ \eqref{laplace gfm} (corresponding to Fig.\ \ref{fig: GFM}) should be modified as
\begin{equation}
  \begin{aligned}
    (\nabla^2 p)_{i,j} & =\frac{p_{i-1,j} -2p_{i,j} +p_{i+1,j}}{\Delta x^2} 
                         -2\frac{[p]_{i,j}}{\Delta x^2} 
                         -\frac{1}{\Delta x} \bigg[\frac{\partial p}{\partial x} \bigg]_{i+1/2,j}
                         +\frac{1}{\Delta x} \bigg[\frac{\partial p}{\partial x} \bigg]_{i-1/2,j} \\
                       & +\frac{p_{i,j-1} -2p_{i,j} +p_{i,j+1}}{\Delta y^2}
                         -\frac{[p]_{i,j-1}}{\Delta y^2},
  \end{aligned}
  \label{laplace gfm mls}
\end{equation}
Similarly, all the jumps should be removed consistently when computing the pressure gradient in the subsequent step. \Ge{The above modification applies to both SLS and MLS, as the compact formulas (Eqs.\ \eqref{gfm+split projection} and \eqref{gfm+split correction}) remain the same; although MLS is clearly more accurate in resolving the near field structure.}


\subsection{Near-field interactions}
\label{subsec: pot}

\Ge{As introduced earlier, colloidal droplets transported in microfluidic devices are subject to various forces, a typical of which is the depletion force. The depletion force arises from the exclusion of the surfactant micelles in the colloidal suspension. It is often characterized as a near-field attracting potential \cite{Asakura_1958, Mewis_colloidal}, and plays a key role in the droplet dynamics \cite{Shen_AS_2016, Shen_thesis}.}
 Below, we first provide a brief background on the colloidal theory of the depletion potential, then present a numerical model to enforce the depletion force using MLS and GFM.


\subsubsection{The colloidal theory of the depletion potential}
\label{depletion theory}

\begin{figure}[t]

 \centering
 \includegraphics[width=.75\columnwidth]{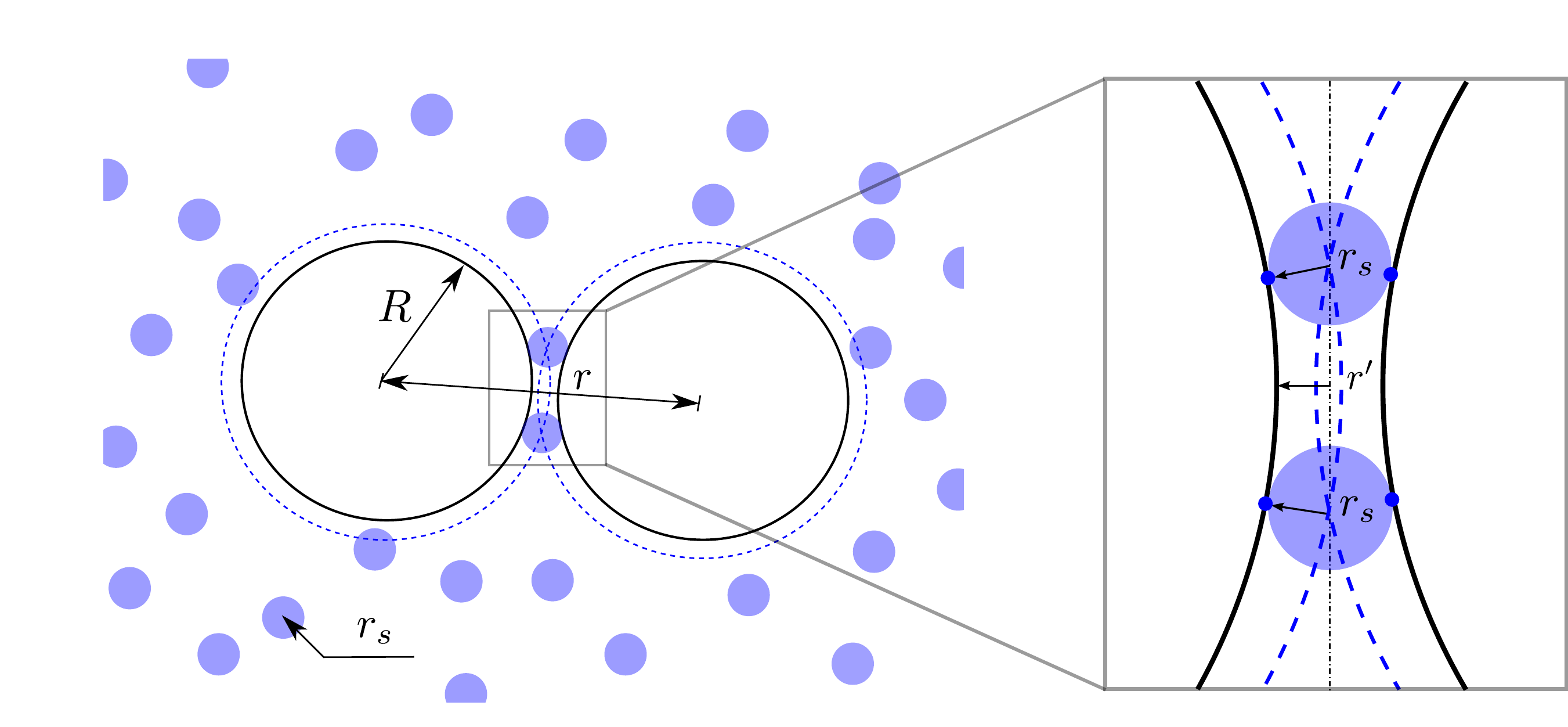}
 \caption{Depletion of surfactant micelles of radius $r_s$ between larger colloidal droplets of radius $R$, separated by distance $r$. The \Ge{dashed lines} around larger spheres represent the region from which the centers of small spheres are excluded. \Ge{They overlap when $r \leqslant 2R+2r_s$.} \Ge{Inset: a zoom-in sketch of two droplets near contact.}}
 \label{fig: overlap}
\end{figure}

The original depletion potential model proposed by Asakura and Oosawa \cite{Asakura_1958} assumes the surfactant micelles as non-interacting hard-spheres. As sketched in Fig.\ \ref{fig: overlap}, a suspension of such small spheres around the large colloidal droplets creates an osmotic pressure on the droplet surface. When the distance between two droplets is less than the diameter of the surfactant micelles, there will be a pressure defect due to the exclusion of the micelles, thus creating an attracting force. Integrating this force with respect to the inter-droplet distance $r$ leads to a potential energy
\begin{equation}
    U(r)=
    \begin{cases}
        \infty & \textrm{if} \quad r \leqslant 2R \\
        - p_{os}V_{ex} & \textrm{if} \quad 2R < r \leqslant 2R+2r_s  \\
        0 \quad & \textrm{otherwise}, \\
    \end{cases}
    \label{pot}
\end{equation}
where $V_{ex}$ is the excluded volume and $p_{os}$ is the osmotic pressure. 
\Ge{For spherical droplets, $V_{ex}$ can be calculated analytically}
\begin{equation}
    V_{ex}(r) = \frac{4\pi (R+r_s)^3}{3}\bigg[ 1- \frac{3r}{4(R+r_s)}+\frac{r^3}{16(R+r_s)^3} \bigg],
    \label{pot V}
\end{equation}
\Ge{where $R$ and $r_s$ are, respectively, the radii of the big and small spheres. The osmotic pressure is given as
\begin{equation}
    p_{os} = nkT,
    \label{pot P}
\end{equation}
where $n$ is the number density of the small spheres, $k$ is the Boltzmann constant, and $T$ is the temperature. The negative sign in Eq.\ \eqref{pot} corresponds to the tendency of the system to reduce its potential energy as the overlap increases. This is equivalent to increasing the total entropy of the small spheres \cite{Melby_PRL_2007}, and it provides a physical description of the depletion force even when the droplets are deformable, or when $p_{os}$ cannot be expressed by the van't Hoff's formula (Eq.\ \eqref{pot P}) \cite{Asakura_1958}.}


\subsubsection{A hydrodynamic model for the depletion force}

\Ge{Based on the above theory, the depletion force acting on a droplet is simply the derivative of the depletion potential, \ie $F(r)=dU/dr=-p_{os}dV_{ex}/dr$. However, $dV_{ex}/dr$ is not always straightforward to evaluate for non-spherical droplets; and unlike rigid-body dynamics, $F(r)$ cannot be applied directly to the motion of a liquid drop.
In order to induce locally an aggregation, we take a closer look at the overlap region. As illustrated in Fig.\ \ref{fig: overlap}, when the surface distance between two colloidal droplets is less than $2r_s$, there is a small area in which the osmotic pressure is subject to a jump. Assuming the concentration of the surfactant micelles changes abruptly, it resembles the jump of the Laplace pressure; however, it will not generate any flow if the pressure is uniform in the depleted region. On the contrary, if the osmotic pressure varies continuously within the overlap, \ie $p'=p'(r')$, then we can write it as a Taylor-series expansion from $r'=r_s$
\begin{equation}
    p'(r'/r_s) = p'(1) + \bigg(\frac{r'}{r_s}-1 \bigg) \frac{\partial{p'}}{\partial{r'/r_s}},
    \label{pot P linear}
\end{equation}
where the distance to the droplet surface $r'$ is normalized by the surfactant micelle radius. An expansion of the osmotic pressure with the distance corresponds to a gradient of the micelle concentration near the gap. And if the micelle is much smaller than the droplet, as it is in many microfluidic devices \cite{Shen_AS_2016}, the gradient will be very sharp. Conversely, when the distance to the surface varies slowly, such as in the gap of a droplet and a flat wall, a uniform pressure will be recovered. Furthermore, a favorable pressure gradient from the overlap center will generate an outflow, pulling the droplets towards each other. Hence, Eq.\ \eqref{pot P linear} provides a hydrodynamic model for the depletion force.}

\Ge{In Eq.\ \eqref{pot P linear}, the gradient of the osmotic pressure $\partial{p'}/\partial{(r'/r_s)}$ is not known \textit{a priori}. It can be obtained by equating the depletion force acting on one droplet, \ie
\begin{equation}
    -p_{os}A_{ex} = \int_\Omega \big(p'(1) - p'(r'/r_s)\big)dS,
    \label{pot P balance}
\end{equation}
where $A_{ex}$ is the effective area of the overlap $\Omega$. Assuming a constant $\partial{p'}/\partial{(r'/r_s)}$, the above yields a linear dependence of the osmotic pressure on $r'$. Note that this is not the same as $p'$ varying linearly with the distance to the overlap center (see Fig.\ \ref{fig: overlap}). A description of the implementation and verification will be shown in the next section.}

\subsubsection{A MLS/GFM-based method for computing the depletion force}

\Ge{
Provided a hydrodynamic model for the depletion force between two droplets, we can easily generalize it to multiple droplets using the MLS.
Thanks to the distance information embedded in the level set functions, it is straightforward to identify the overlap region of arbitrary geometries. Furthermore, as the jump of the osmotic pressure occurs only across the overlap shell, we can define
\begin{equation}
    [p']_\Omega = p'(r'/r_s) - p'(1),
    \label{pot P jump}
\end{equation}
similar to the Laplace pressure jump $[p]_\Gamma$ implemented by the GFM. Based on these observations, we propose a numerical method to compute the depletion force as laid out in Algorithm \ref{al: depletion}.}

\begin{algorithm}[t]
 Enter the pressure solver. Compute the right-hand side of Eq.\ \eqref{gfm+split projection}.\\ 
 \For{$m=1:(N-1)$}{
  Get the level set for droplet m, $\phi_m$.\\  
  \For{$n=(m+1):N$}{
   Get the level set for droplet n, $\phi_n$.\\   
    \textbf{where} $\phi_m<r_{s}$ and $\phi_n<r_{s}$ \textbf{do} $r'=(\phi_m+\phi_n)/2$, tag as \textit{overlap}.\\
    Compute $[p']_\Omega$ from Eqs.\ \eqref{pot P balance} and \eqref{pot P jump} within \textit{overlap}.\\
    \ForAll{$i,j,k$}{
     \eIf{entering overlap}{
     Add the osmotic pressure jump $[p']_{i,j,k}$.\\
     }{
     Remove the osmotic pressure jump $[p']_{i,j,k}$.\\
     }
    }
  }
 }
 Solve for $p^{n+1}$ regularly using the FastP* and GFM. Exit the pressure solver.\\
 \caption{A pseudo code for computing the depletion force.}
 \label{al: depletion}
\end{algorithm}

The overall idea of Algorithm \ref{al: depletion} is to enforce the depletion attraction in the projection step through the use of MLS and GFM. \Ge{Specifically, we first locate the overlap region of a pair of droplets with its own level set function, and define $r'$ as the average of the two distances. Then, Eq.\ \eqref{pot P balance} can be integrated numerically to obtain $\partial{p'}/\partial{(r'/r_s)}$, which together with Eqs.\ \eqref{pot P linear} and \eqref{pot P jump} gives $[p']_\Omega$. This variable pressure jump manifests itself as a modification term on the right-hand side of Eq.\ \eqref{gfm+split projection}, allowing us to use GFM to impose it across a sharp overlap shell. The resulting flow is divergence-free provided that all the jump terms are removed consistently in the correction step. Therefore, Eqs.\ \eqref{gfm+split projection} and \eqref{gfm+split correction} are re-formulated as 
\footnote[1]{\Ge{Eqs.\ \eqref{gfm+split correction} and \eqref{complete correction} are identical in form; however, $[p']_\Omega$ has to be removed when evaluating $\nabla_g p^{n+1}$ and $\nabla_g \hat{p}$ in Eq.\ \eqref{complete correction}, as it is done in Eq.\ \eqref{gfm grad}}}
\begin{equation}
    \nabla ^2 p^{n+1} = \nabla_g^2 ([p]_\Gamma+[p']_\Omega) 
    + \nabla \cdot \bigg[ \big(1-\frac{\rho_0}{\rho^{n+1}}) \nabla_g \hat{p} \bigg] + \frac{\rho_0}{\Delta t} \nabla \cdot {\bm u}^*,
  \label{complete projection}
\end{equation}}
and 
\begin{equation}
    {\bm u}^{n+1} = {\bm u}^* -\Delta t \bigg[\frac{1}{\rho_0} \nabla_g p^{n+1} + \big(\frac{1}{\rho^{n+1}} - \frac{1}{\rho_0}\big)\nabla_g \hat{p} \bigg].
  \label{complete correction}
\end{equation}

\paragraph{Approaching drops} 

\Ge{We verify the depletion force model and its numerical implementation by simulating 2 to 14 approaching droplets in a quiescent fluid environment. Specifically, we set the droplet radius $R=0.5$, the computational domain $3\times3\times3$, and the resolution $\Delta x=1/32$. The radius of the surfactant micelle is set to be $r_s=1/16$, corresponding to $2\Delta x$. The viscosity and density ratios of the droplet to the ambient fluid are both 1. The non-dimensional parameters are 
$La=2000$
and $Fr=\infty$, leading to a reference Laplace pressure jump $p_\sigma=80$ and neglected gravity. The uniform osmotic pressure is either $10$ or $40$.}

\begin{figure}[t]
 \begin{center}
  \includegraphics[width=\columnwidth]{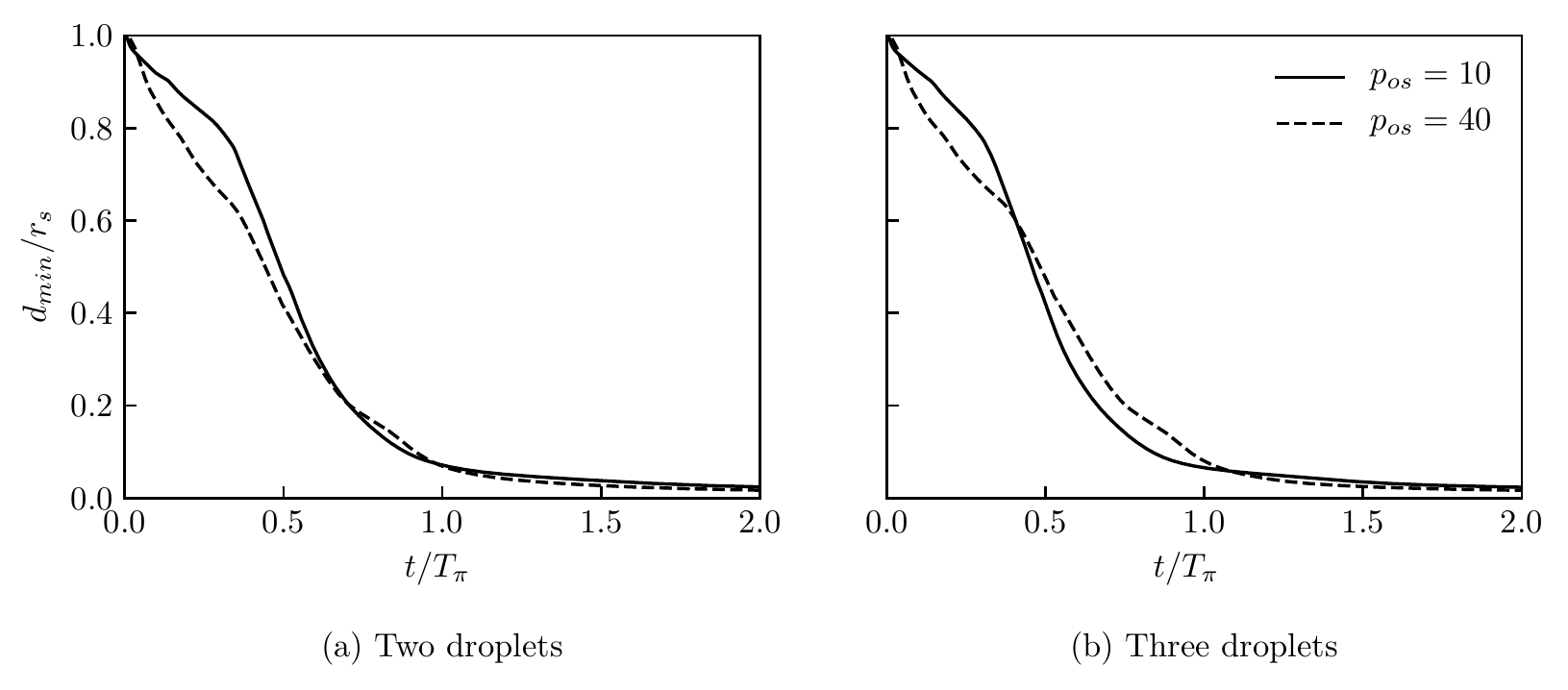}

  \begin{picture}(0,0)
   \put(-70,80){\includegraphics[height=2cm]{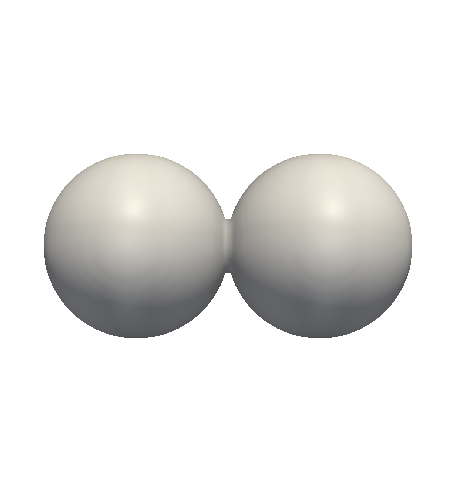}}
   \put(150,80){\includegraphics[height=2cm]{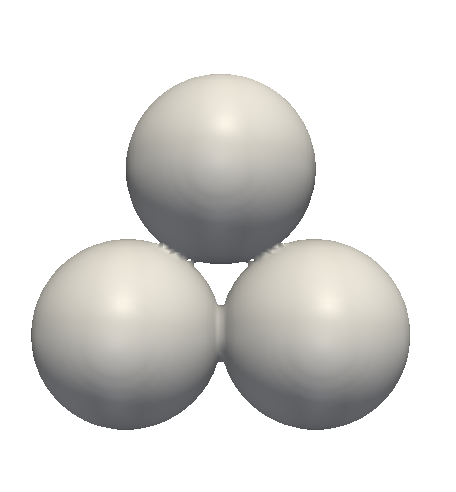}}
  \end{picture}

 \end{center}
 \caption{Minimal distance between the droplet surfaces as function of time in the presence of depletion forces proportional to $p_{os}=10$ (solid line) and $p_{os}=40$ (dashed line). Simulation of (a) two droplets and (b) three droplets suspended in an initially quiescent fluid. Due to symmetry, only the minimal distance is plotted.}
 \label{fig: dep demo}
\end{figure}

\begin{figure}[t]
\centering
  \subfigure[Diamond]{\includegraphics[width=.2\columnwidth]{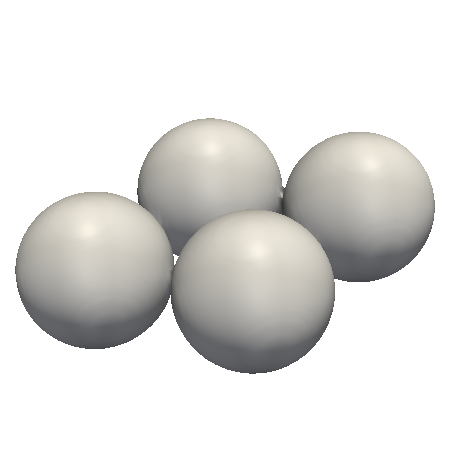}}
  \subfigure[Tetrahedron]{\includegraphics[width=.2\columnwidth]{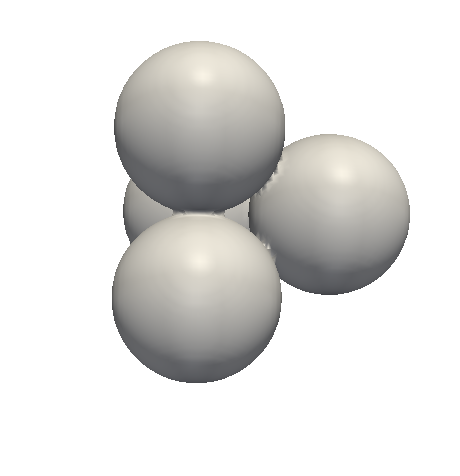}}
  \subfigure[Quintuplet]{\includegraphics[width=.2\columnwidth]{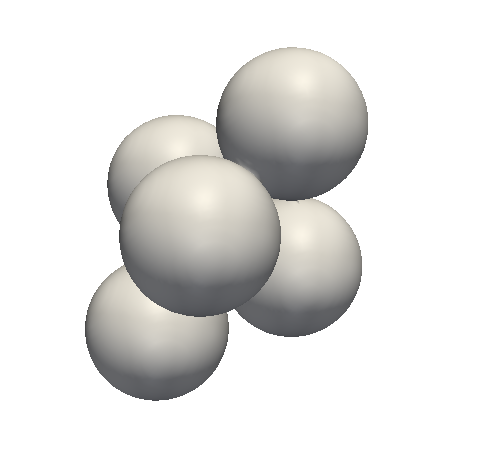}}
  \subfigure[FCC]{\includegraphics[width=.2\columnwidth]{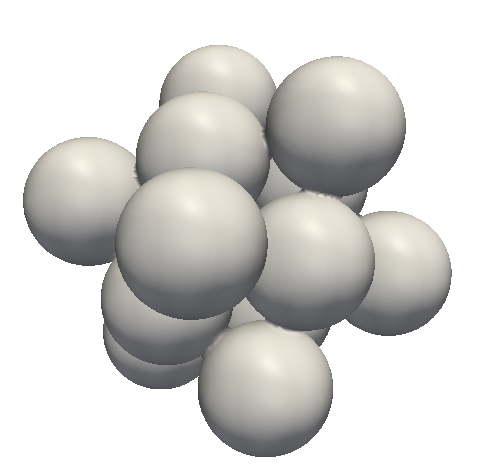}}
  \caption{\Ge{Examples of droplet clusters of different structures.}}
  \label{fig: clusters}
\end{figure}

\Ge{The temporal evolutions of the minimal surface distances in the case of two and three droplets are shown in Fig.\ \ref{fig: dep demo}. Here, time is scaled by a factor $T_\pi=(r_s/R)(p_\sigma/p_{os})$. The droplets, originally separated by a distance of $r_s$, get closer to the limit of the grid spacing at $t \approx T_\pi$. For the present study, we let the droplets aggregate without applying any repulsion models, except that the magnitude of the osmotic pressure is reduced when $d_{min}/r_s<0.1$. The smooth approaching in all cases and the collapse of the distance curve clearly evidence an attracting depletion force. To assess the robustness of the method, we further tested clustering of droplets into shapes from a 2D diamond to a face-centered cubic (FCC) composed of 14 drops, illustrated here in Fig.\ \ref{fig: clusters}. FCC represents the unit structure of one of the most compact sphere packings. Therefore, we can conclude that the hydrodynamic model implemented by the MLS/GFM-based method is accurate and robust in computing the depletion forces.}

\section{Conclusion}
\label{sec: conclusion}

A numerical method mainly intended for the hydrodynamic simulations of colloidal droplets in microfluidic devices has been developed and validated. The code is based on \Ge{an efficient and sharp} solver of the incompressible, two-fluid Navier-Stokes equations, and uses a mass-conserving level set method to capture the fluid interface. This combination provides a general framework for any multiphase flow problems (see \eg our recent study on jet instabilities \cite{Loiseau_PRF_2016}), and allows us to develop specific methods for the simulations of droplets in saturated surfactant suspensions with depletion forces as in the recent experiment in \cite{Shen_AS_2016}. Particularly, we have developed or extended four numerical techniques to improve the general accuracy:

\begin{enumerate}

    \item A mass-conserving, interface-correction level set method (ICLS) is proposed. As a standalone level set module, it is efficient, accurate, guarantees global mass conservation, and is simple to implement. It also enables corrections that can depend on the local curvature or any other parameter of interest. 
        
    \item A \Ge{geometric} estimation of the interface curvature based on nodal curvatures is introduced. As an important ingredient both for the mass correction (ICLS) and the surface tension computation, we show that the calculation converges in second-order both in 2D and 3D, and can lead to machine-zero spurious currents for a stationary 2D droplet.
    
    \item The ghost fluid method (GFM) for the computation of surface tension is combined with the FastP* method \cite{Dodd_JCP_2014}. This enables the use of \Ge{FFT-based solvers for a direct pressure solve}, and can accurately account for surface tension at large density ratios.
    
    \item A ghost fluid/multiple level set (GFM/MLS-based) method is also proposed to compute the \Ge{interaction force} caused by depletion potentials between multiple droplets or between droplets and a nearby wall. The approach can possibly be extended to account for surfactant diffusion at the interface and in the liquid.

\end{enumerate}

The last technique applies specifically to the simulation of colloidal droplets in microfluidic devices. This will enable us to further explore the effects of the near-field interactions as those observed experimentally in  \cite{Shen_AS_2016}, and potentially improve the design of microfluidic devices.
In addition, the combination of the GFM for sharp interfaces and the FastP* method \cite{{Dodd_JCP_2014}} can be exploited for the simulations of droplet in turbulent flows as in \cite{Dodd_JFM_2016}, adding an accurate representation of evaporation thanks to the ICLS approach proposed here.


\section*{Acknowledgments}

The work is supported by the Microflusa project. This effort receives funding from the European Union Horizon 2020 research and innovation programme under Grant Agreement No.\ 664823. O.T. acknowledges support from Swedish Research Council (VR) through the Grant Nr.\ 2013-5789. L.B. and J.-C.L. also acknowledge financial support by the European Research Council grant, no.\ ERC-2013-CoG-616186, TRITOS. The computer time was provided by SNIC (Swedish National Infrastructure for Computing).
Last but not least, Z.G. thanks Mehdi Niazi, Michael Dodd, Walter Fornari, Dr.\ Olivier Desjardins, Dr.\ S\'{e}bastien Tanguy, Dr.\ Marcus Herrmann, and Dr.\ David Salac for interesting and helpful discussions.


\appendix

\section{\Ge{Discretization error of $\int_\Gamma {\bm n}\cdot{\bm u}_c d\Gamma$}}
\label{appendix-dis_err}
\Ge{
Similar to \cite{Engquist_JCP_2005}, we define the discretization error
\begin{equation}
    E= \bigg| \bigg( \prod_{k=1}^{d} \Delta x_{k} \bigg) \sum_{j \in Z^d} \hat{\delta}_\epsilon (\Gamma,g,{\bm x}_j)
    -\int_\Gamma {\bm n}\cdot{\bm u}_c d\Gamma \bigg|,
  \label{app-def}
\end{equation}
where $\hat{\delta}_\epsilon$ is a Dirac delta function of variable strength $g$ supported on the surface $\Gamma$, and ${\bm x} \in \mathbb{R}^d$. Following the derivations in Sec.\ \ref{subsec: ICLS}, the extension of $g$ to $\mathbb{R}^d$ is provided by Eq.\ \eqref{correction vel sol}, allowing one to write
\begin{equation}
    E= \bigg|\bigg( \prod_{k=1}^{d} \Delta x_{k} \bigg) 
    \sum_{j \in Z^d} \frac{\delta V}{\delta t} \frac{f_s \delta_\epsilon(\phi({\bm x}_j))|\nabla \phi({\bm x}_j)|}{A_f} 
    -\int_\Gamma {\bm n}\cdot{\bm u}_c d\Gamma \bigg|.
  \label{app-step1}
\end{equation}
Here, $\delta_\epsilon(\phi)$ is a one dimensional regularized delta function depending on the level set $\phi$, and the expression is simplified noting that $\bm{n} \cdot \nabla{\phi} = |\nabla \phi|$ (it does not have to be a distance function). By definition, $A_f=\int_\Gamma f_s \delta_\epsilon (\phi) |\nabla \phi| \,d\Gamma$, discretely reducing Eq.\ \eqref{app-step1} to
\begin{equation}
    E= \bigg| \frac{\delta V}{\delta t}
    -\int_\Gamma {\bm n}\cdot{\bm u}_c d\Gamma \bigg|.
  \label{app-step2}
\end{equation}
Comparing with Eq.\ \eqref{vol change}, it is obvious that $E=0$. That is, the discretization error of $\int_\Gamma {\bm n}\cdot{\bm u}_c d\Gamma$ used in the mass correction is identically zero, independent of the choice of the regularized delta function.
}




\bibliographystyle{elsarticle-num}

\bibliography{1}

\end{document}